\documentclass[aps,prl,twocolumn,superscriptaddress]{revtex4-1}

\usepackage{graphicx,hyperref,bbm,amssymb,amsfonts,amsmath,times,subfigure,footnote}

\newcommand{\sx}[1]{s^x_{#1}}
\newcommand{\sy}[1]{s^y_{#1}}
\newcommand{\sz}[1]{s^z_{#1}}
\def\ii{{\rm i}}
\newcommand{\ket}[1]{|{#1}\rangle}
\newcommand{\bra}[1]{\langle {#1}|}
\newcommand{\braket}[2]{\langle{#1}|{#2}\rangle}
\newcommand{\tr}{\mathrm{tr}}
\newcommand{\ave}[1]{\langle{#1}\rangle}

\date{\today}

\begin{document}

\title{Diffusive and subdiffusive spin transport in the ergodic phase of a many-body localizable system}
\author{Marko \v Znidari\v c}
\affiliation{Physics Department, Faculty of Mathematics and Physics, University of Ljubljana, 1000 Ljubljana, Slovenia}
\author{Antonello Scardicchio}
\affiliation{Abdus Salam ICTP, Strada Costiera 11, 34151 Trieste, Italy}
\affiliation{INFN, Sezione di Trieste, Via Valerio 2, 34127 Trieste, Italy}
\author{Vipin Kerala Varma}
\affiliation{Abdus Salam ICTP, Strada Costiera 11, 34151 Trieste, Italy}
\begin{abstract}
We study high temperature spin transport in a disordered Heisenberg chain in the ergodic regime. 
By employing a density matrix renormalization group technique for the study of the stationary states of the boundary-driven Lindblad equation we are able to study extremely large systems ($400$ spins).
We find both a diffusive and a subdiffusive phase depending on the strength of the disorder and on the anisotropy parameter of the Heisenberg chain. Studying finite-size effects we show numerically and theoretically that a 
very large crossover length exists that controls the passage of a clean-system dominated dynamics to one observed in the thermodynamic limit. 
Such a large length scale, being larger than the sizes studied before, explains previous conflicting results. We also predict spatial profiles of magnetization in steady states of generic nondiffusive systems.
\end{abstract}

\maketitle

\textit{Introduction.---} There are ever increasing technological capabilities in simulating isolated quantum systems through cold atomic gases \cite{coldatoms} and, recently, 
through coupled, controlled superconducting qubits \cite{annealers}. 
While there is a commensurately good theoretical handle on capturing ground state properties of such systems \cite{Sachdev11}, understanding their 
dynamical properties, especially away from the ground state, is fraught with analytical and numerical challenges.

Despite this, in the recent years we have witnessed a change in paradigm in the study of isolated quantum systems, in particular with regard to the role that disorder plays in such systems. 
The turning point came about from the study of Anderson localization \cite{Anderson58} in interacting, many-body quantum systems \cite{BAA}. 
The observation that disorder and quantum effects can hinder transport 
(of energy, charge or spin) even at an infinite temperature and in the presence of interactions \cite{Pal10} opened the door to a new phenomenology of a so-called many-body localized (MBL) phase exhibiting many unique and interesting properties. 
Slow growth of entanglement \cite{Znidaric08,Bardarson12}, emergent integrability \cite{IOMs}, protection of symmetries \cite{Chandran14a}, and change in the properties of eigenstates 
\cite{DeLuca13,Luitz15,Goold15} are a few of the peculiar properties of this newly identified phase; 
see review~\cite{rew} for a comprehensive list. The implications of the new MBL physics, being inherently robust, are far reaching, going from fundamental physics to the theory of 
quantum computation \cite{Altshuler10}, some of which have already been experimentally probed~\cite{exper}.

While the deep MBL region (in one dimensional systems) is well understood, much remains to be said about the conducting regime and the transition to it.
Although both aspects are important, here we focus on characterizing the conducting phase, in particular its transport properties, in, what is by now, an archetypal model that harbors the MBL phase, i.e., the one dimensional anisotropic Heisenberg model.

Generic arguments and numerical evidence on very small systems (about 20 spins) have been put forward for the existence of subdiffusive transport of spin~\cite{Agarwal,Luitz15a,Potter15,Sarang15} and 
energy \cite{Potter15, Vosk14, Vipin15}. 
A number of recent works have analyzed its spin transport properties in 
the ergodic phase, finding different results. Applications of numerical renormalization group recipes~\cite{Vosk14,Potter15} (which should be valid in some region preceding the MBL transition) find a 
subdiffusive phase for energy and spin transport, 
with continuously changing subdiffusion exponents. Numerical calculations (on small systems) find either (i) a subdiffusive regime close to the MBL 
transition preceded, at smaller disorder, by a transition to diffusion~\cite{Agarwal}, or 
(ii) a subdiffusive regime all the way to zero disorder without a sharp transition~ in between~\cite{Luitz15a}.

\begin{figure}[bbp]
\centerline{\includegraphics[width=3.2in]{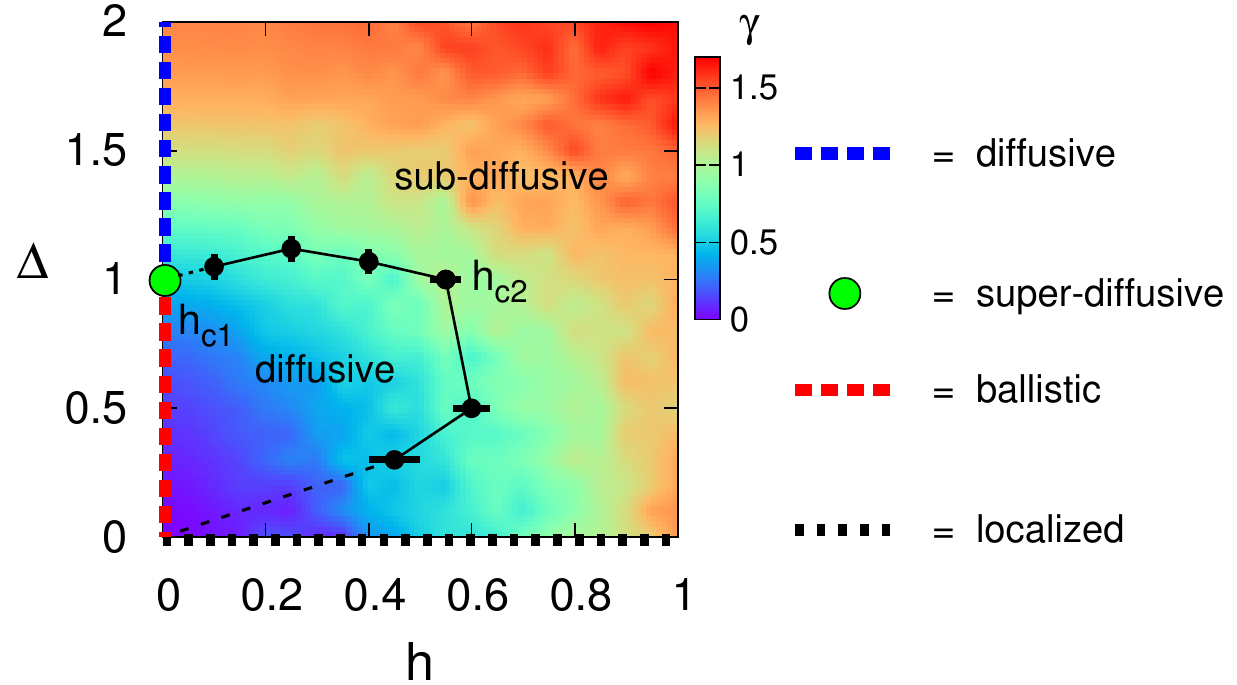}}
  \vspace*{-0.15cm}
\caption{(Color online) 
Phase diagram of a disordered anisotropic Heisenberg model in the high temperature ergodic phase. 
At a critical disorder strength $h_{c2}$ there is a transition from diffusive 
to subdiffusive spin transport. 
Black circles with error bars denote $h_{c2}$ determined from the steady-state current scaling $j \sim 1/L^\gamma$ in large systems ($L \approx 400$ sites, see Fig. \ref{fig:1}). The underlying 
colors are for illustrative purposes and denote $\gamma$ obtained from small 
systems $L\le 7$, which, nevertheless, correctly depicts the two regimes, except close to $h=0$ and $\Delta=0$.
}
\label{fig:0}
\end{figure}
In this Letter we resolve the issue by clearly demonstrating the existence of both a \emph{diffusive} and 
a \emph{subdiffusive} regime within the ergodic phase; in so doing we identify the transition point (its value being different from previous claims), 
and provide an explanation for the finite-size effects in terms of a mechanism for equilibration of the conserved quantities of the integrable clean model~\cite{Bethe31}. 
Which regime occurs depends both on the anisotropy parameter of the clean system and on the disorder strength. The phase diagram summarizing these points is shown in Fig. \ref{fig:0}. 

We study spin transport by coupling the system to a combination of injecting and absorbing reservoirs using the Lindblad equation~\cite{Lindblad}, see e.g.\ Refs~\cite{Hartman03,JSTAT09,Clark13}. 
Using time-dependent density matrix renormalization group (t-DMRG) to study nonequilibrium steady 
state (NESS), in particular the scaling of spin current with system size, we 
are able to reach considerably larger systems of up to $L=400$ spins, 
allowing us to study all the length scales involved in the problem. 
We find that a crossover scale $L_*$ determines up to which size the system will 
behave like its clean counterpart, while for $L \gg L_*$ the disorder becomes 
relevant and the thermodynamic limit (TDL) is achieved. 
We present a theoretical explanation of $L_*$ based on a weak-disorder 
perturbation theory for the quasiparticles of the clean, integrable model, and find that it grows quickly when $h\to 0$. Because $L_*$ can be one order of 
magnitude larger than systems explored in previous studies, this lends an explanation to the contradictory nature of previous results. Therefore one finds good cause to pause before taking 
their conclusions at face value (we remark that in classical 
transport studies~\cite{review-class} it has been observed in numerous systems that large sizes are indeed needed before the TDL is reached).

{\em Model.---} The Hamiltonian of a disordered anisotropic Heisenberg chain is 
$H_{\rm XXZ}:=\sum_{k=1}^{L-1} \sx{k}\sx{k+1}+\sy{k}\sy{k+1}+\Delta \sz{k}\sz{k+1}+\frac{h_k}{2}\sz{k}+\frac{h_{k+1}}{2}\sz{k+1},$ 
where $s_k^{\alpha}=\frac{1}{2}\sigma_k^\alpha$ are spin-$1/2$ operators ($\sigma_k^\alpha$ are Pauli matrices), 
$\Delta$ is the anisotropy, and $h_k \in [-h,h]$ are uniform independent disorder fields of maximum strength $|h|$ at site $k$. 

For $h=0$ the model is the famous (clean) XXZ model, finding use in many areas of physics. While originally constructed as a toy model to explain ferromagnetism~\cite{Bethe31}, today its 
main attractiveness comes due to its integrability by the Bethe ansatz. It is a canonical example of a nontrivial (i.e., non-quadratic) quantum integrable model. 
Not least, the XXX spin chain is realized with high accuracy also in many real materials~\cite{materials}. Transport studies of the clean XXZ model have a long history and despite its integrability, even today, not many rigorous results are available. Limiting our discussion to high-temperature transport results, of interest to present work, 
one exception is a rigorous bound~\cite{Prosen11} on the ballistic transport for $\Delta < 1$. 
At the isotropic point $\Delta=1$ numerical results show~\cite{PRL11,JSTAT11} superdiffusive transport 
(faster than diffusive, but slower than ballistic), supported also by classical correlations~\cite{Robin12}. 
An independent indication for superdiffusive transport at $\Delta=1$ is also arrived at from observing the behavior for slightly smaller or larger $\Delta$. 
In the limit $\Delta \rightarrow 1^-$ there is a prevailing opinion~\cite{D0} that at $\Delta=1$ the Drude weight is zero 
(i.e. slower than ballistic transport). On the other hand, in the gapped phase $\Delta>1$ numerics indicates diffusive transport, with the diffusion constant rapidly increasing (diverging)~\cite{PRL11,diff} 
as $\Delta \rightarrow 1^+$.
 
Nonzero disorder $h$ breaks integrability, with even less reliable transport results existing. In this Letter we shall focus on the regime below the MBL transition point which occurs at 
$h_{c3}(\Delta=1)\approx 3.7$~\cite{Pal10,Luitz15}. Our goal is to mimic, through numerical simulation, what an experimentalist would do to measure transport: 
we couple the system at its two ends to ``magnetization'' reservoirs that induce a NESS carrying spin current. Concretely, we use the 
Lindblad master equation~\cite{Lindblad} describing Markovian evolution of the system's density matrix,
\begin{equation}
{{\rm d}}\rho/{{\rm d}t}=\ii [ \rho,H_{\rm XXZ} ]+ \frac{1}{4}\sum_{k=1}^4 \left( [ L_k \rho,L_k^\dagger ]+[ L_k,\rho L_k^{\dagger} ] \right),
\label{eq:Lin}
\end{equation}
where Lindblad operators $L_k$ effectively account for generic magnetization driving by two ``baths'', and are $L_1=\sqrt{1+\mu}\,\sigma^+_1, L_2= \sqrt{1-\mu}\, \sigma^-_1$ at the left end, 
and $L_3 =  \sqrt{1-\mu}\,\sigma^+_L, L_4= \sqrt{1+\mu}\, \sigma^-_L$ at the 
right end, $\sigma^\pm_k=(\sigma^{x}_k \pm {\rm i}\, \sigma^{y}_k)/2$. Provided there is an asymmetry in driving between the two ends, i.e., $\mu \ne 0$, a 
nonzero steady-state current is induced. We remark that, while a microscopic derivation~\cite{Breuer} of such a driving might be difficult in a 
condensed-matter context, 
our approach is rather pragmatic: in a generic nonintegrable system such as ours details of a boundary driving should not matter for the bulk physics. 
Also, at long times in NESS possible non-Markovian effects should not be important. Therefore, the results that we obtain for the bulk are independent of the details of the driving.

For our choice (\ref{eq:Lin}) the NESS $\rho_\infty$ is always unique and 
therefore any initial state $\rho(0)$ eventually converges to $\rho_\infty$, 
$\lim_{t \to \infty} \rho(t)=\rho_\infty$. For zero (equilibrium) driving, $\mu=0$, the steady state is a trivial infinite temperature state $\rho_\infty(\mu=0) \sim \mathbbm{1}$. 
We will always use small driving $\mu=0.001$, meaning that our $\rho_\infty$ 
is always close to the identity, in other words, we are in a linear response regime (see Ref.~\cite{Supp} for data) and at infinite temperature. 
Current of a conserved quantity is defined by a commutator with a local Hamiltonian density (such that the continuity equation holds, $\dot{s}_k^{z}=j_k-j_{k-1}$), which for 
the spin current leads to $j_k:=\sx{k}\sy{k+1}-\sy{k}\sx{k+1}$. Our 
central quantity is the expectation value of $j_k$ in the NESS, $\tr{(j_k \rho_\infty)}$, 
which is, due to stationarity, also independent of site index $k$ and will 
be denoted simply by $j$. The Lindblad driving that we use is in a way the 
simplest one that will induce a nonzero spin current while at the same time the disorder-averaged energy current is zero. Therefore, we are able 
to focus exclusively on spin transport. Note that by an antisymmetric 
disorder with $h_k=-h_{L+1-k}$ we can achieve that the NESS energy current 
is zero for each disorder 
realization, which though leads to the same results for large $L$~\cite{Supp}.
\begin{figure}[ttp]
\centerline{\includegraphics[width=3.3in]{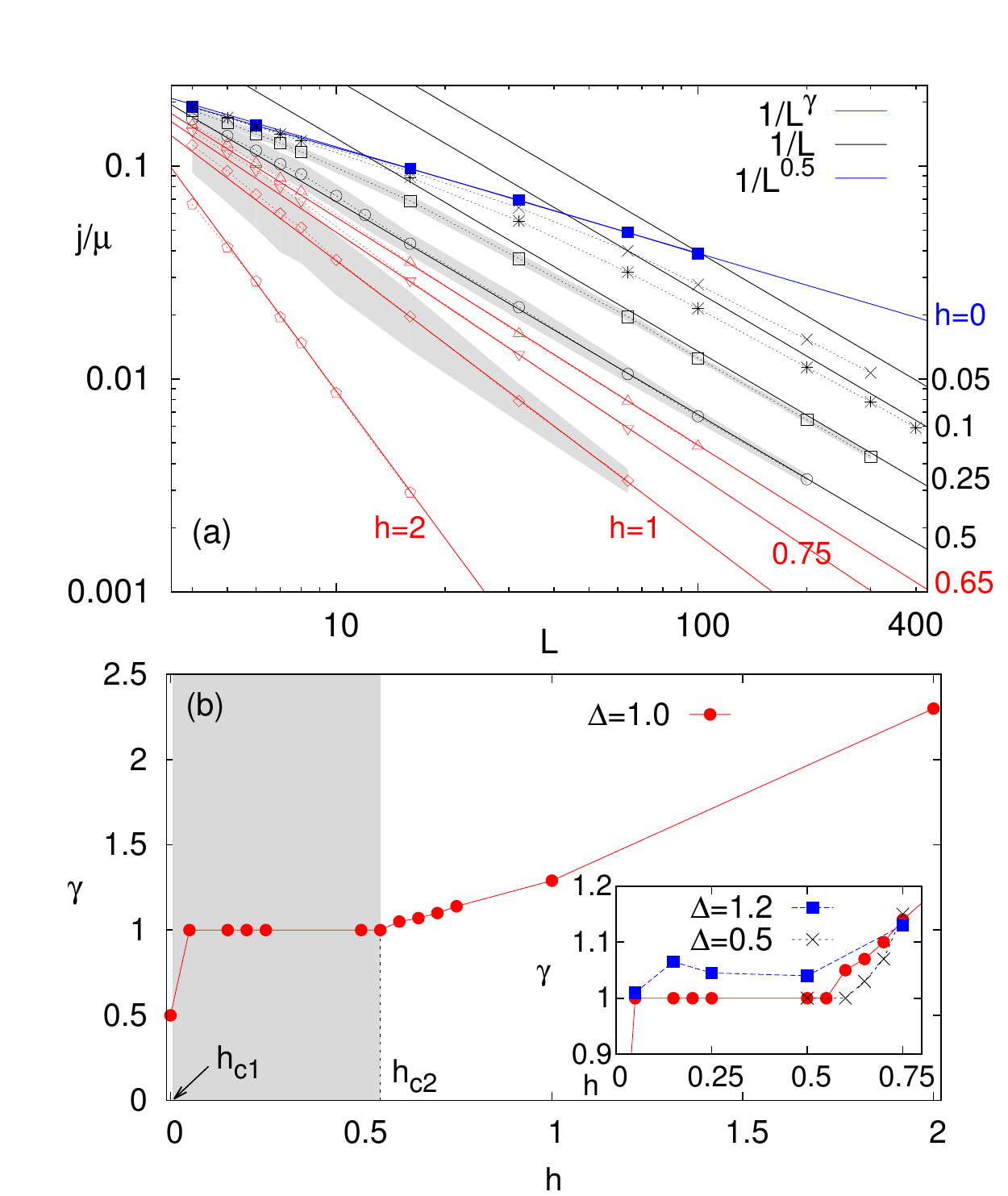}}
\caption{(Color online) 
Spin transport in the ergodic phase of the Heisenberg model.
(a) Scaling of the average NESS spin current $j$ with system size, 
$j \sim 1/L^\gamma$, for $\Delta=1$. Points are numerical data, lines are 
best fitting $1/L^\gamma$, with $\gamma=1$ for $0<h<h_{c2}\approx 0.55$ (black), and $\gamma>1$ for $h_{c2} < h < h_{c3}$ (red). For $h=1$, $h=0.5$ and $h=0.25$ 
the gray shading denotes standard deviation $\sigma(j)$ of current distribution (for $h=0.25$ it is barely visible as it is smaller than the size of square 
points). (b) Dependence of $\gamma$ on the disorder strength. At a critical disorder strength $h_{c2}$ one gets a transition from diffusive to subdiffusive spin 
transport, while at $h_{c1}=0$ there is a discontinuous transition from 
superdiffusive (for $h=0$) to diffusive for $h_{c1} < h < h_{c2}$. 
Inset: similar data for $\Delta=1.2$ and $\Delta=0.5$.}
\label{fig:1}
\end{figure}

{\em Current scaling.---} For each disorder realization we solve the Lindblad 
equation for the NESS $\rho_\infty$ using a t-DMRG method, simulating time evolution $\rho(t)$ until the state converges to $\rho_\infty$ (for $L \le 8$ we 
also used exact diagonalization). We can reach systems with up to $L=400$ sites~\cite{foot1}, thereby revealing new interesting physics. Details of our 
t-DMRG implementation can be found in Ref.~\cite{JSTAT09}; for numerical parameters see Ref.~\cite{Supp}. 

We perform ensemble averaging of the NESS spin current $j$ to obtain the average current, which is our main quantity of interest. 
The disorder sample size $M$ is chosen for each $h$ and $\Delta$ such that the estimated statistical uncertainty $\sigma(j)/\sqrt{M}$, where $\sigma(j)$ is standard deviation of the NESS current, 
is $\approx 2\%$ or less. For an example of $\sigma(j)$ see Fig.~\ref{fig:1} . We have also studied the whole NESS spin current probability distribution $p(j)$ for our 
disorder ensemble, finding that for small $h$ (e.g., $h=0.5$) it is well described by a Gaussian, while at larger $h$ (e.g., $h=2$) it is clearly 
non-Gaussian, though being well described by a log-normal distribution. We also observe~\cite{Supp} that away from the MBL 
transition relative current fluctuations $\sigma(j)/j$ go to zero in the TDL, as expected for an  
ergodic phase. 

For $h< h_{c3}$ we expect the average current to scale as $j \sim  1/L^\gamma$ (in the MBL phase $h>h_{c3}$ one would 
instead have $j \sim \exp{(-\kappa L)}$), which is indeed borne out by numerical results. We recall that $\gamma=1$ signifies a diffusive transport 
(and validity of a phenomenological transport law $j = -D \nabla s^{z}$, where $D$ is a diffusion constant), while $\gamma>1$ is called subdiffusive 
($\gamma \to \infty$ signifying localization, e.g., for $h\ge h_{c3}$), and 
$\gamma<1$ is a superdiffusive transport ($\gamma=0$ being ballistic). All other scaling exponents can be expressed in terms of $\gamma$, provided there is 
only one scaling exponent. Scaling $j \sim 1/L^\gamma$ implies that a finite-size diffusion constant goes as $D \sim L^{1-\gamma}$, while the current 
autocorrelation function decays as $C_{jj}=\langle j(t)j(0)\rangle \sim 1/t^{\eta}$ with $\eta=2\gamma/(1+\gamma)$. 
Assuming the variance of initial inhomogeneities to grow as $\langle x^2 \rangle_{c} \sim t^{2\beta}$, meaning that an excitation needs time $t \sim L^{1/\beta}$ to traverse the system, 
at fixed excitation density the current will scale as $j \sim L/t$, resulting in the relation $\beta=1/(\gamma+1)$, which has been observed in a number of classical 
systems~\cite{Li03}. Spin autocorrelation function in turn scales as $C_{zz}(t) \sim 1/t^\beta$ at long times and, using the continuity equation in momentum space, the 
low-frequency conductivity will in turn scale as $\sigma(\omega) \sim \omega^\alpha$ with $\alpha=(\gamma-1)/(\gamma+1)$.

From data for the average current (Fig.~\ref{fig:1}(a)) we can extract 
the scaling exponent $\gamma$ and plot it as a function of $h$ (Fig.~\ref{fig:1}(b)). At the isotropic point $\Delta=1$ (i) we find a transition from 
subdiffusive (for $h> h_{c2}$) to diffusive transport at $h_{c2} \approx 0.55$ (for more precise data see Ref.~\cite{Supp}); and (ii) there 
is another transition at $h_{c1}=0$ at which spin transport in the TDL goes discontinuously from diffusive to superdiffusive $\gamma=0.5$~\cite{PRL11,JSTAT11}.

Repeating the analysis for $\Delta \ne 1$~\cite{Supp} 
we find at $\Delta=1.2$ and the smallest $h=0.05$ considered that $\gamma=1.01 \pm 0.01$, and therefore determine $h_{c2}(\Delta=1.2) \lesssim 0.05$. 
On the other hand, varying $\Delta$ at fixed $h=0.1,0.25,0.4$, we find transition points at critical $\Delta$ equal to $\approx 1.05, 1.12, 1.07$, respectively, decreasing as $h \to 0$. 
We therefore conclude that the phase line likely connects to the point $\Delta=1$, $h=0$ (see Fig.~\ref{fig:0}). 
In the gapped phase $\Delta > 1$, where the clean model displays spin diffusion at high temperature (although higher current moments seem to have a nondiffusive scaling~\cite{fluctXXZ}), 
a very weak disorder suffices for the onset of subdiffusion. For $\Delta =0.5$ the transition point is $h_{c2}(\Delta=0.5) \approx 0.60$, while for $\Delta=0.3$ it 
is $h_{c2}(\Delta=0.3) \approx 0.45$, and therefore decreases for weak interactions. 
The limit of small interactions $\Delta \rightarrow 0^+$, where one approaches a singular Anderson regime in which \textit{both} diffusion and subdiffusion cease \cite{Anderson58}, is rather interesting. 
The scattering length due to small interactions scales as $\xi_\Delta \sim 1/\Delta^2$, while the Anderson localization length is $\xi_A \sim 1/h^2$. 
Therefore, a necessary condition to see diffusion is $\xi_\Delta \lesssim \xi_A$, i.e., $\Delta \gtrsim h$ (dashed line in Fig.~\ref{fig:0}). 
Note that the $\Delta \rightarrow 0^+$ limit and the TDL do not commute.

In our simulations we also obtain NESS magnetization profiles $\ave{s^z_k}$. We constructed a heuristic theory, accounting for length-dependent diffusion constants when $\gamma \neq 1$, explaining the 
observed magnetization profiles, even for finite $L$ that are not yet in the TDL; for details see Ref.~\cite{Supp}.

{\em Weak disorder.---} Here we shall be interested in the regime $h<h_{c2}$. 
We have seen in Fig.~\ref{fig:1} that the transition from subdiffusive to 
diffusive spin transport happens at a relatively small disorder strength. In 
addition, for even smaller $h \ll h_{c2}$ the asymptotic scaling $j \sim 1/L$ is reached only at a sufficiently large $L \gg L_*$, with $L_*$ increasing 
with decreasing $h$. 
For instance, at $h=0.1$ even $L=300$ is not yet completely in the asymptotic 
diffusive regime. Therefore, for small $h$ there is a nontrivial 
characteristic length-scale $L_*$ below which 
transport will appear to be superdiffusive (similar to the clean 
anomalous $j \sim 1/L^{0.5}$ scaling) while for $L>L_*$ one eventually 
starts to see diffusion.

\begin{figure}[ht!]
\centerline{\includegraphics[width=3.3in]{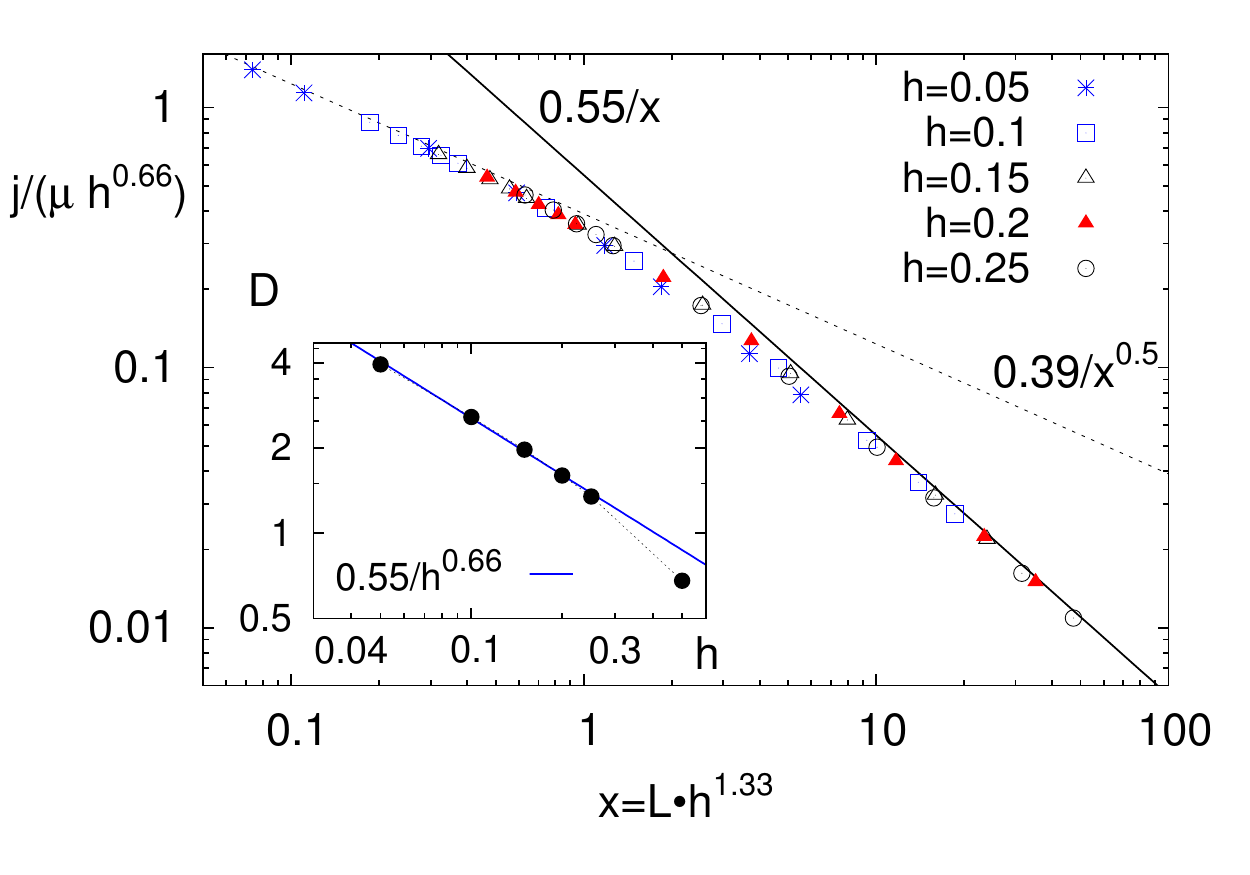}}
\vspace{-0.5cm}
\caption{(Color online) Scaling collapse of NESS current with system size for small 
$h$ and $\Delta=1$ (additional data to that in Fig.~\ref{fig:1}(a) are plotted). In the main plot we see that for $L > L_* \sim 1/h^{1.33}$ the scaling is 
diffusive, $j \sim 1/L$, while at shorter lengths it is asymptotically 
superdiffusive, $j \sim 1/L^{0.5}$, 
the same as in the clean model. Data for $L \le 300$ are shown ($L\le 400$ for $h=0.1$). The inset shows the scaling of the diffusion constant $D$ with $h$, 
with the straight (blue) line suggesting $D \sim 1/h^{0.66}$ divergence for 
small $h$.}
\label{fig:smallh}
\end{figure}

In Fig.~\ref{fig:smallh} we show data for $\Delta=1$ and small $h\le 0.25$, scaling the 
horizontal axis (system size) by $L_* \sim 1/h^\nu$, that is using a 
scaling variable $x:=Lh^\nu$. In addition scaling the vertical axis 
by $h^{\nu-\delta}$ we can achieve a collapse of all points on a single 
scaling curve $j \sim h^{\nu-\delta} f(Lh^\nu)$, with the best empirical 
scaling exponents being $\nu \approx 1.33$ and $\delta \approx 0.66$. 
Because diffusive transport implies $f(x) \sim 1/x$, the diffusion constant 
diverges at small disorder as $D \sim 1/h^\delta$. We note that $\delta$ 
and $\nu$ are not independent: for small values of disorder 
one should recover an $h$-independent behavior of the clean isotropic model 
$j \sim 1/\sqrt{L}=h^{\nu/2}/\sqrt{x}$, leading to $\nu/2=\nu-\delta$. Importantly, one 
can see from Fig.~\ref{fig:smallh} that the asymptotic diffusive spin transport is observed clearly only for $L \gtrsim 20/h^{1.33}$ (equal to $\approx 45$ sites even at $h=h_{c2}$), explaining why 
previous studies (limited to $L \lesssim 30$) either could not see 
diffusion at small $h$, or made an incorrect prediction for $h_{c2}$. 
In the inset of Fig.~\ref{fig:smallh} we show the scaling of the diffusion 
constant with $h$ as determined by independently fitting $j(L)$ with $j \sim D/L$, obtaining the same $\delta=0.66 \pm 0.1$. 

Let us now theoretically explain the obtained scaling exponent $\nu$. 
At small disorder we can consider the excitations of the clean, integrable 
system, as almost freely propagating, except for a few scattering over 
the disorder. The motion of these excitations can be summarized in a law of 
the form $x\sim t^{\beta}$, where $\beta=1$ for $\Delta<1$, while $\beta=2/3$ for $\Delta=1$. 
One can then ask what is the typical length a particle of energy $\epsilon$ 
has to go before it changes its momentum $k$ due to disorder. For the time $\tau_k$ 
from the first collision we can use Fermi's golden rule, $\frac{1}{\tau_k}=2\pi\sum_{k'}|\bra{k'}V\ket{k}|^2\delta(\epsilon-\epsilon_{k'})$, 
where $V=\sum_{j=1}^Lh_j\ket{j}\bra{j}$, $\epsilon_k=2J\cos(k)$ for the case 
of free quasiparticles, and $\braket{k}{j}=e^{\ii kj}/L$. With these expressions, averaging over disorder 
(the quantity is self-averaging in the large $L$ limit anyway), and passing 
from the sum to the integral we find $\frac{1}{\tau_k}=\frac{h^2}{24J}\frac{1}{\sqrt{1-\epsilon^2/(2J)^2}}$. In the middle of the spectrum $\epsilon=0$, and so we have $\tau\sim J/h^2$. 
Then, using $x \sim t^\beta$, we can predict that $L_* \sim 1/h^{2\beta}$, and taking the anomalous superdiffusion exponent $\beta=\frac{2}{3}$, 
one obtains $\nu=2\beta=\frac{4}{3}\approx 1.33$. In Fig.~\ref{fig:smallh} 
we can see that the optimal exponents with which we scaled both axes agree within numerical precision with the predicted $\nu\approx 1.33$ and 
$\nu-\delta\approx 0.66$.

The above argument can also be used to predict the scaling exponent $\nu$ 
for $\Delta <1$, where again the clean transport is faster than the 
diffusive one induced by disorder. Taking ballistic $\beta=1$ results in $\nu=2$, i.e., $L_* \sim 1/h^2$ for small $h$. This prediction is confirmed by numerics~\cite{Supp}. Understandably, compared to the isotropic 
model, adding disorder to a ballistic model one will need larger systems 
to eventually see diffusion, or, equivalently, for a system of fixed 
length $L$ larger disorder is needed to bring in diffusion. We also 
numerically determined the exponent $\delta$ of the diffusion constant's 
divergence, obtaining $\delta=1.4\pm 0.1$ for $\Delta=0.8$, 
while $\delta=1.8\pm 0.2$ at $\Delta=0.5$~\cite{Supp}. 
For $\Delta>1$ physics of $L_*$ is different because one has a 
competition of two equally fast (diffusive) transport channels, a scattering due to interaction in a clean system and a scattering due to disorder. For sufficiently small disorder the clean diffusive mode always dominates, leading to $\delta=\nu=0$~\cite{Supp}.

\textit{Conclusions.---} We studied the nonequilibrium steady-state spin 
current at infinite temperature in the disordered Heisenberg chain with 
boundary drives. The fact that we are able to simulate transport dynamics 
in system sizes up to $L=400$ unveiled a critical length scale $L_*$ in 
the ergodic phase of the many-body localizable spin chain, above which the 
disorder acts as a relevant perturbation to the clean 
integrable limit; for $L < L_*$ the system ``pretends'' to be the clean 
system in its transport dynamics. In particular, at the isotropic point a 
finite critical disorder strength separates the diffusive and subdiffusive 
regimes of spin transport. In the gapped phase $\Delta>1$ we find that 
this critical disorder strength rapidly decreases, making a diffusive system 
very unstable to disorder, immediately leading to subdiffusion. We may 
understand breaking of integrability upon the introduction of disorder 
primarily as being due to scattering of excitations rather than dephasing, 
explaining the obtained dynamical scaling exponents. We also propose a 
shape of the steady-state magnetization profiles in nondiffusive systems. 
Our approach using the time-dependent density matrix renormalization group 
unveils a more comprehensive methodology to studying transport properties 
in disordered systems, offering an exciting tool to study transport of 
many quantities and different models.

We would like to thank J.~Goold and, in particular, David Huse for discussions. M.\v Z. acknowledges Grants No. J1-7279, No. N1-0025, and No. P1-0044 from the Slovenian Research Agency.

\newpage
\clearpage
\section*{Supplemental Material}

\subsection{Numerical method}

To simulate $\rho(t)$ we use a 4th order Trotter-Suzuki decomposition of the Liouville propagator ${\rm e}^{{\cal L}t}$ into small time-steps of length $\Delta t=0.2$, see Ref~\cite{JSTAT09} 
for details of our open-systems implementation as well as for references about the t-DMRG method in a non-dissipative setting. 
For simulations of NESSs there are two main parameters that determine the efficiency of the method. One is the required dimension $\chi$ of matrices in the matrix product operator (MPO) ansatz used to 
describe $\rho(t)$ -- computational complexity of one time-step scales as ${\cal O}(L\chi^3)$ -- while the other is the Liouvillian gap that determines relaxation rate (speed with which $\rho(t)$ converges 
to $\rho_\infty$). In our simulations we typically relaxed for time $\sim 150$ for small sizes of $L=16$, while for larger systems relaxation times could be $\sim 4000$. 
Typical MPO sizes used were $\chi=40-60$ with which we could get a few percent accuracy in the current. 
In particular cases where higher accuracy was sought, e.g., $h=0.05$ for $\Delta=1.2$ (see Fig.~\ref{fig:delta}(a)), we used up to $\chi=150$.

\subsection{Current distribution}

\begin{figure}[b!]
\centerline{\includegraphics[width=3.3in]{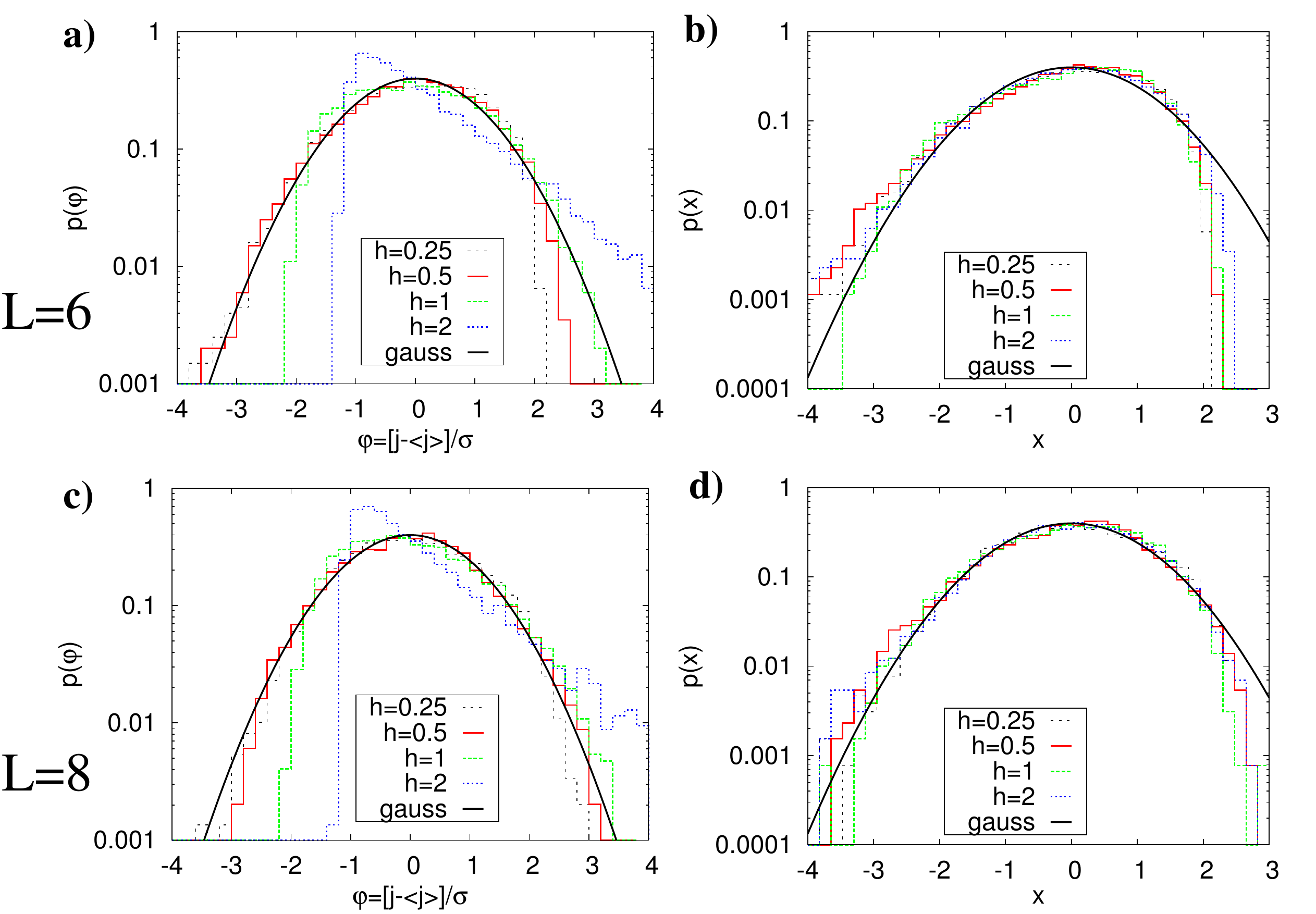}}
\caption{(Color online) Steady-state current distribution for $L=6$ (top row, (a) and (b)) and $L=8$ (bottom row, (c) and (d)). Left plots show distribution of scaled current 
$\varphi:=\frac{j-\bar{j}}{\sigma(j)}$, while the right ones show distribution of scaled logarithm of currents $x:=\frac{\log_{10}{j}-\overline{\log_{10}{j}}}{\sigma(\log_{10}{j})}$. 
Full black curves are Gaussian distributions, and all data correspond to $\Delta=1$.}
\label{fig:podj}
\end{figure}

We studied distribution of NESS spin current $p(j)$ for our disorder ensemble. In Fig.~\ref{fig:podj} we show such distribution for small sizes $L=6$ and $L=8$ for which we are able to gather enough statistics ($10000$ disorder realizations for $L=6$ and $7400$ for $L=8$).
We can see that, while distribution of current changes considerably with $h$ (left plots), the distribution of the log-current seems to be rather stable and described by a normal distribution, i.e., distribution of current is log-normal. Small deviations from the log-normal distribution visible in right plots in Fig.~\ref{fig:podj} are likely due to finite-size effects (they decrease going from $L=6$ to $L=8$). 

\begin{figure}[t!]
\centerline{\includegraphics[width=3.in]{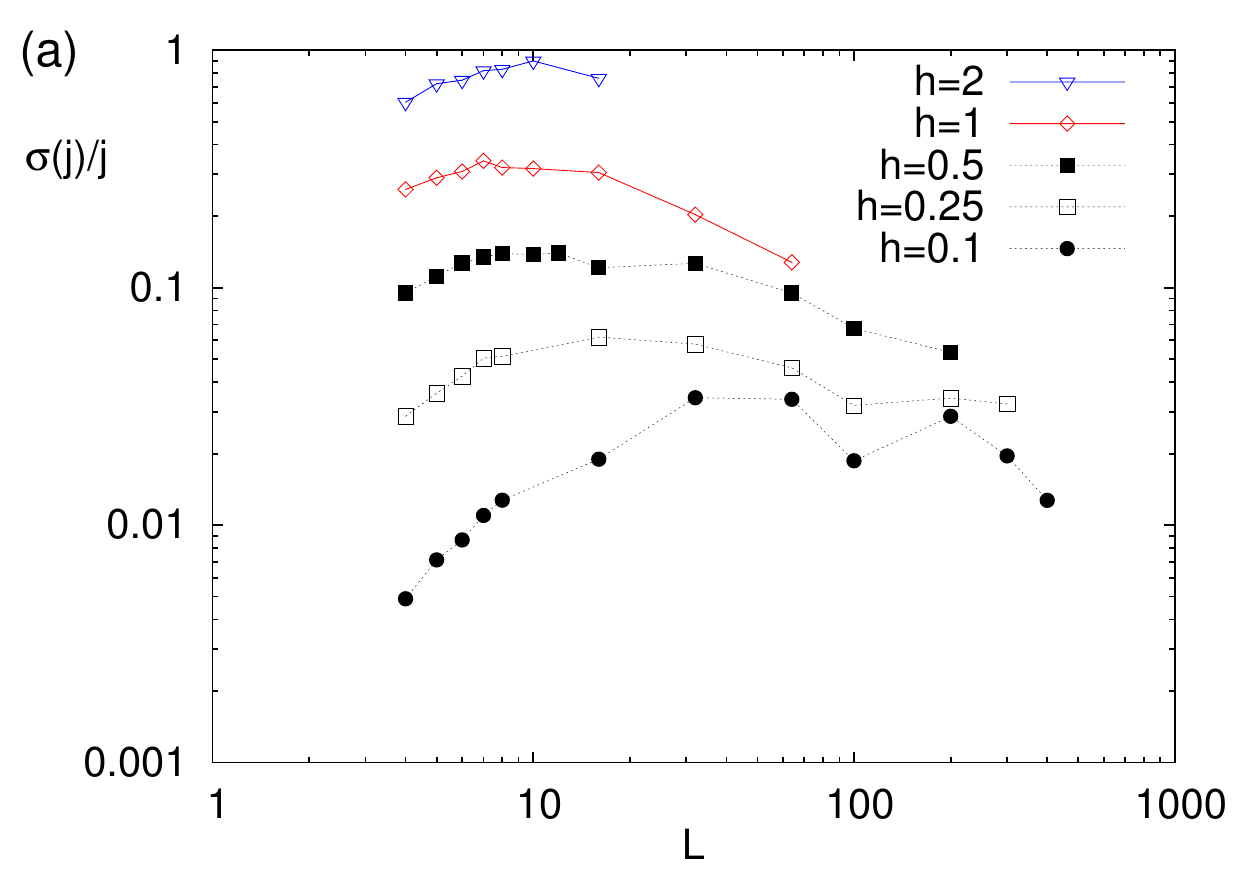}}
\centerline{\includegraphics[width=1.65in]{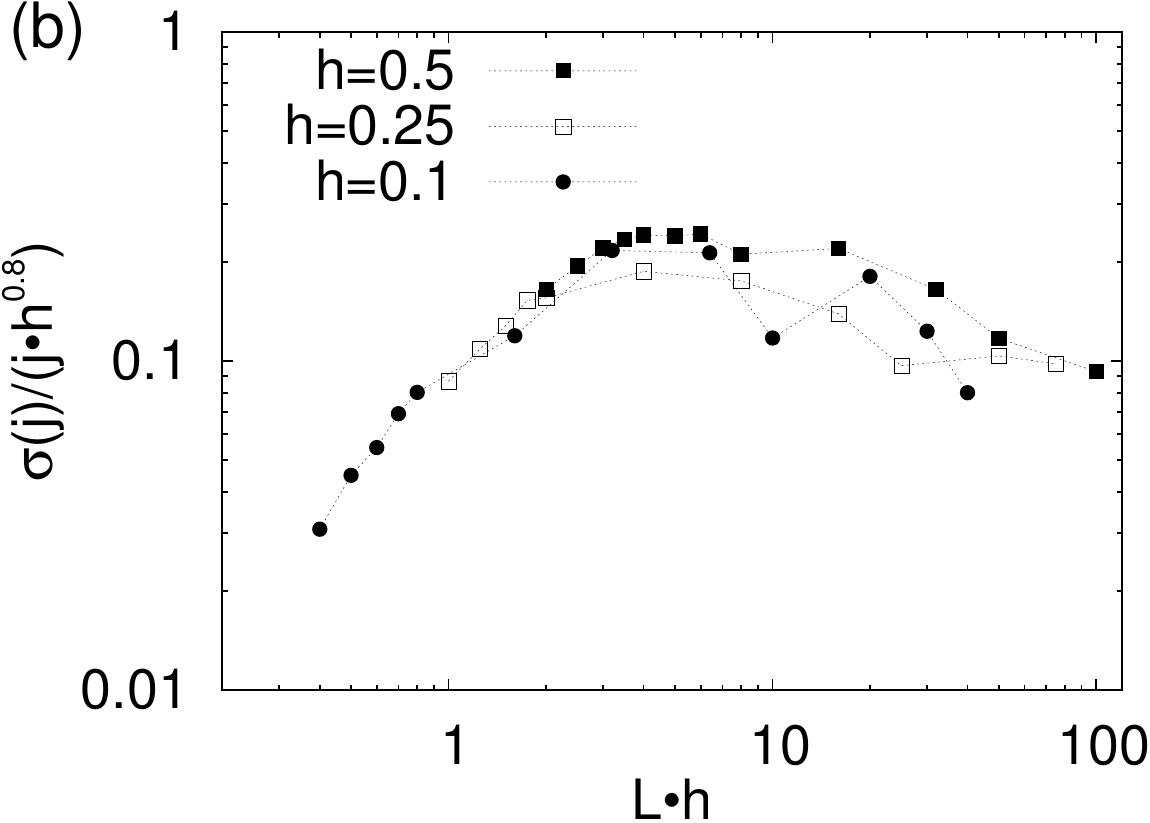}\includegraphics[width=1.65in]{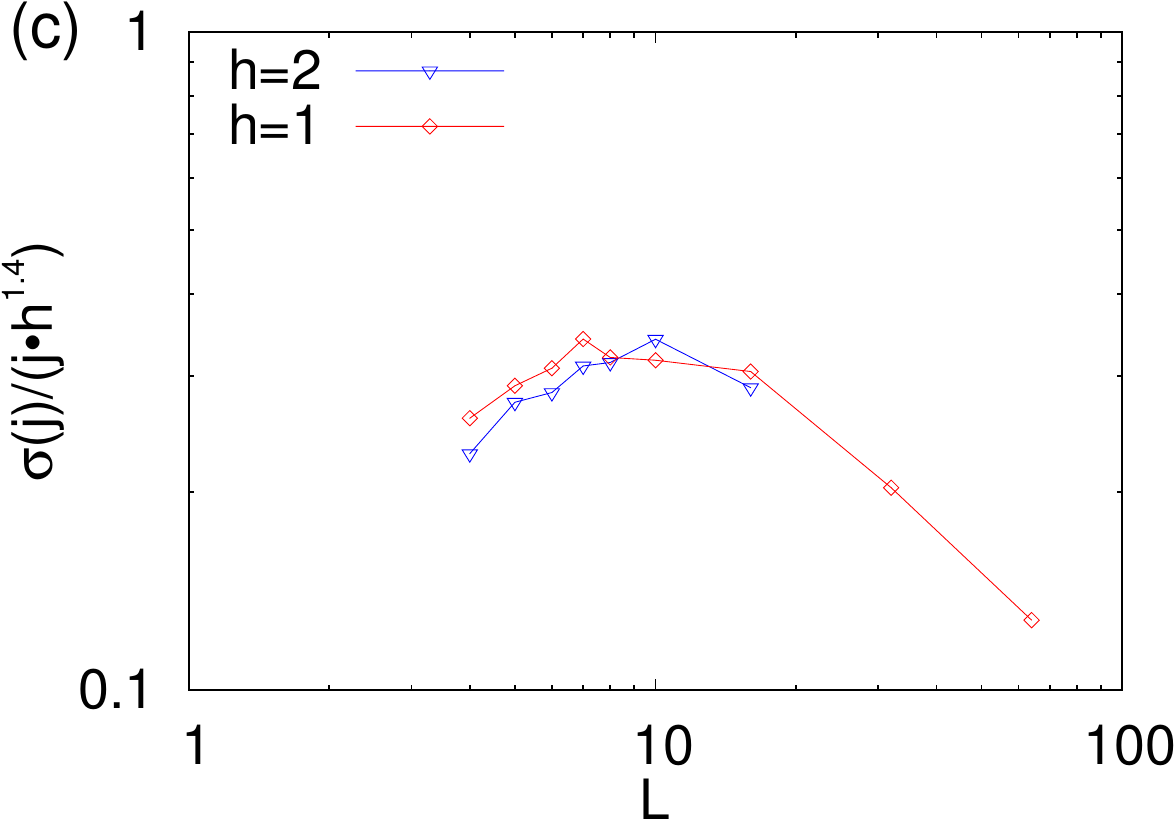}}
\caption{(Color online) (a) Relative NESS current fluctuations $\sigma(j)/j$ for $\Delta=1.0$. For all disorder strengths away from the MBL transition $h_{c3}$ standard deviation of current distribution $\sigma(j)$ divided by the average current $j$ goes to zero in the TDL. Large fluctuations in curves $\sigma(j)/j$ for small $h$ are due to small sample size, e.g., just $M=10$ disorder realizations for $h=0.1$ and $L=100$. (b) and (c) show the same data as (a) but with scaled axes; (b) for $h< h_{c2}$, and (c) for $h> h_{c2}$.}
\label{fig:sigmaj}
\end{figure}

In Fig.~\ref{fig:sigmaj} we show dependence of the standard deviation $\sigma(j)$ of the steady-state current distribution $p(j)$, observing that relative fluctuations go to zero in the TDL, 
regardless of whether one has diffusive or subdiffusive transport. Note that in the MBL phase, $h> h_{c3}$, relative fluctuations will not go to zero in the TDL (data not shown). Observe also that at the transition point $h_{c2}$ nonequilibrium steady-state fluctuations $\sigma(j)$ per se do not exhibit any particular divergence.

We used $\sigma(j)$ to infer the necessary disorder sample size $M$ over which we had to average in order to get statistical uncertainty $\sigma(j)/\sqrt{M}$ in the average $j$ to be of order $2\%$. Typically this required 
few hundred realization for $h=2$, while e.g. only $M=2$ have been used for $h=0.1$ and largest $L=400$ because $\sigma(j)/j$ is already of order $1\%$ (see Fig.\ref{fig:sigmaj}(a)).

\subsection{Critical disorder}

\begin{figure}[t!]
\centering
  \begin{tabular}{@{}cccc@{}}
    \hspace*{-0.5cm}\raisebox{-3ex}{\includegraphics[origin=c, scale=0.62]{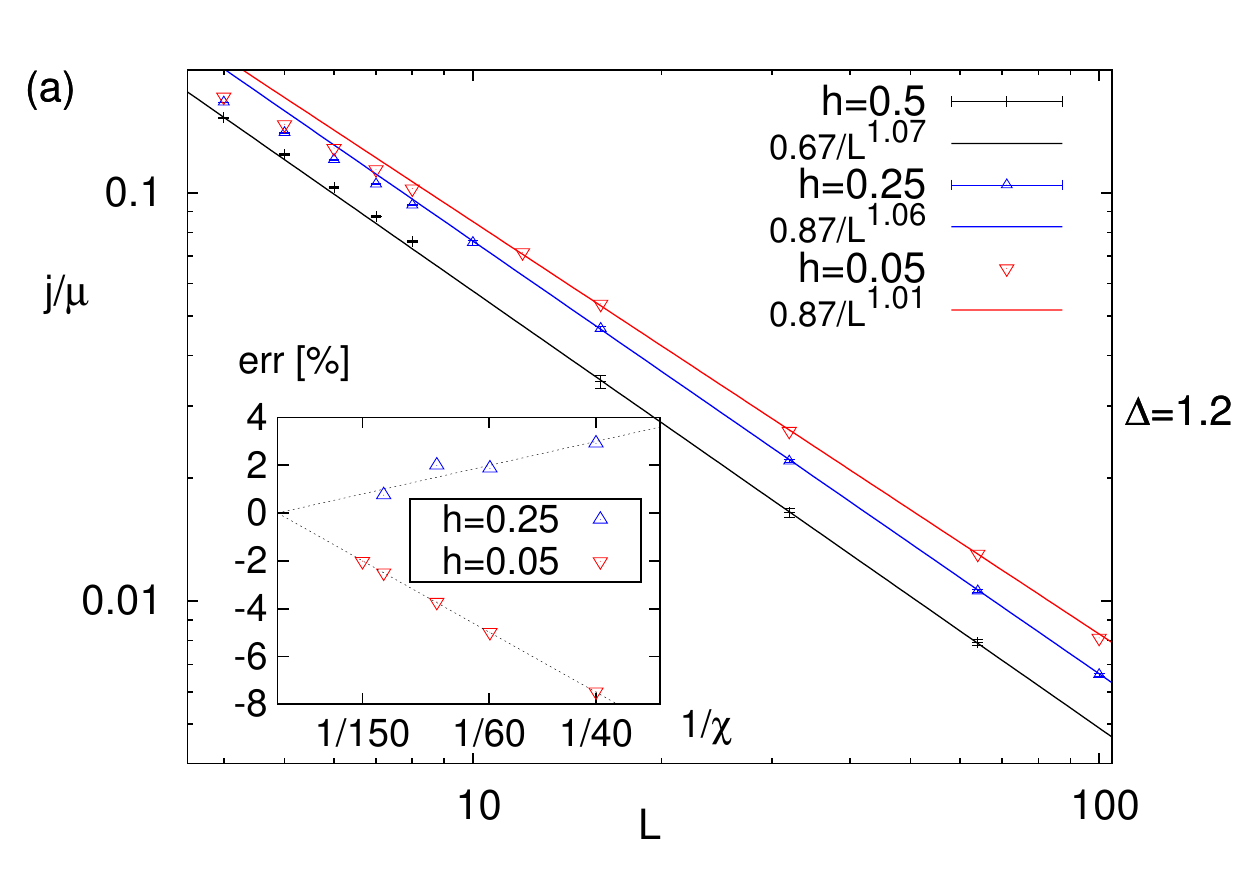}} \\
    \hspace*{-0.5cm}\raisebox{-3ex}{\includegraphics[scale=0.62]{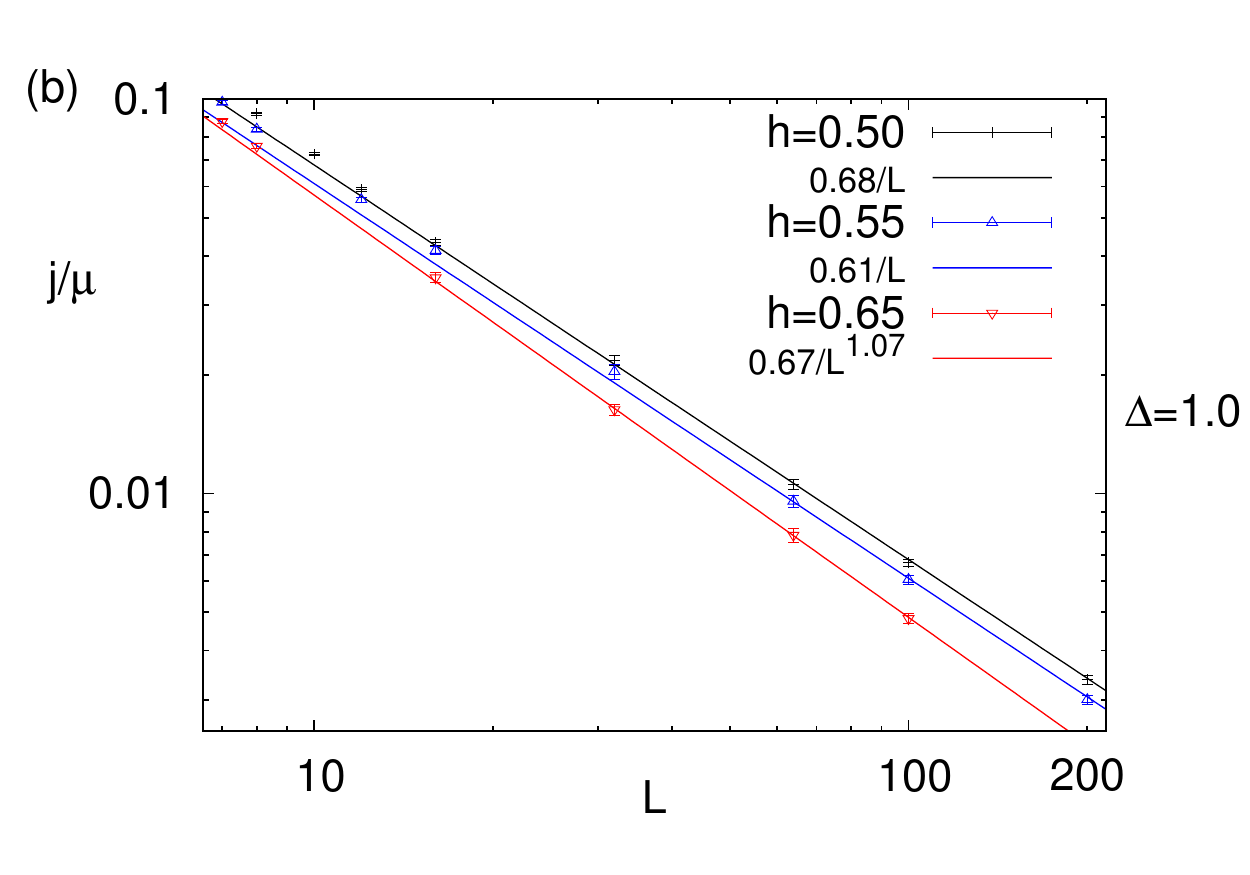}} \\
    \hspace*{-0.5cm}\raisebox{-3ex}{\includegraphics[scale=0.62]{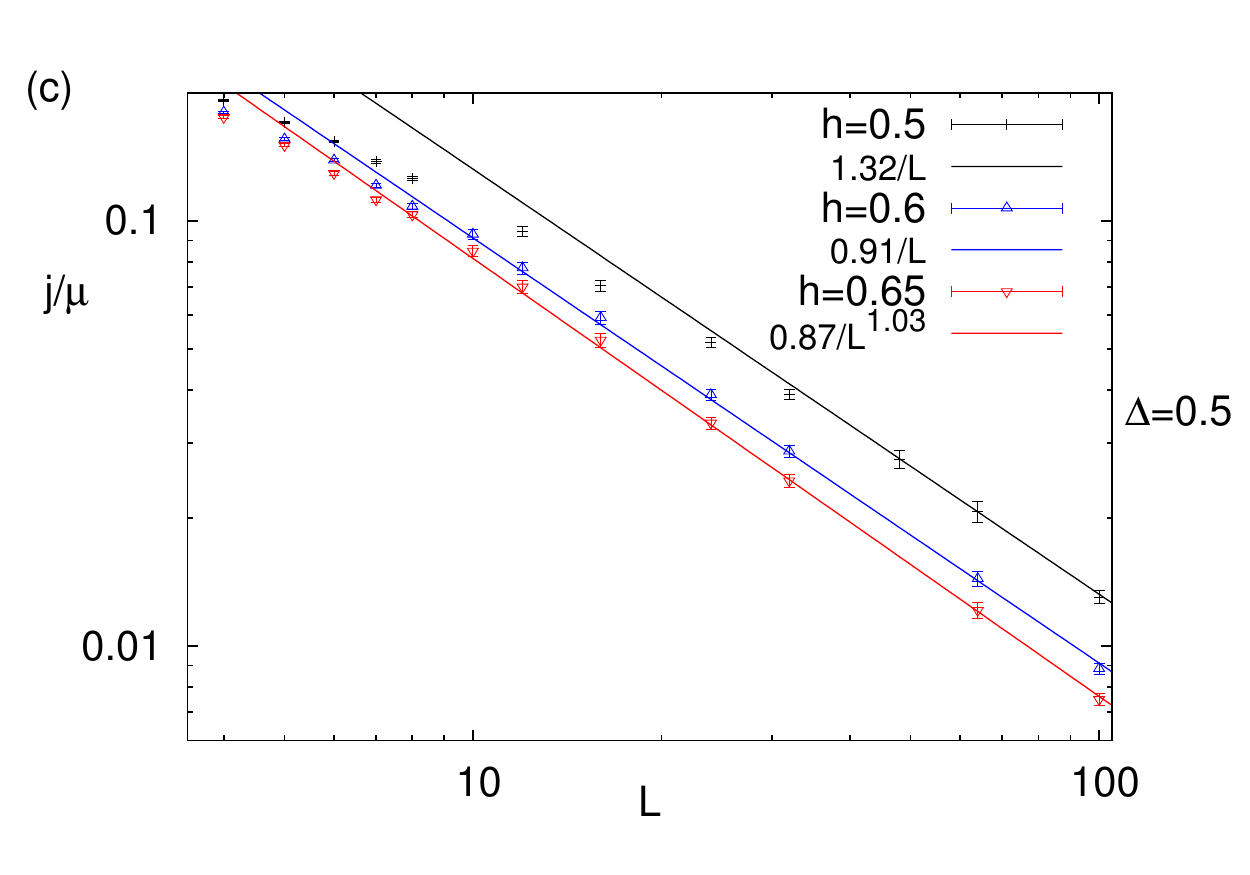}}
  \end{tabular}
  \vspace*{-0.35cm}
\caption{(Color online) Determining $h_{c2}$ for $\Delta=1.2, 1.0, 0.5$, from top to bottom. In all plots we show numerical (points) data together with error bars and a best fitting dependence $j \sim 1/L^\gamma$ (lines).
(a) $\Delta=1.2$. For $h=0.5, 0.25$ we show the average current while for $h=0.05$ a single disorder realization suffices (sample-to-sample variation is less than $0.5\%$). 
Inset: convergence of accuracy with the MPO dimension $\chi$, $err:=[j(\chi)-j(\infty)]/j(\infty)\cdot[100\%]$. (b) $\Delta=1.0$. We show data for $h=0.5, 0.55$ and $h=0.65$, and can see that within the accuracy of about $1\%$ for $h\le h_{c2}=0.55$ the scaling exponent is $\gamma=1.00$, i.e., diffusive spin transport. Observe that the asymptotics is approached only for $L \gtrsim 30$. (c) $\Delta=0.5$. Compared to $\Delta=1.0$ the transition can be estimated to happen at slightly larger $h_{c2} \approx 0.6$.
}
\label{fig:delta}
\end{figure}
For $\Delta=1.2$ and $h=0.05$, where we sought very high accuracy in order to determine whether $\gamma >1$, 
we used extrapolation in the MPO dimension $\chi$ in order to determine the current. We found that dependence of current for finite $\chi$ nicely fits 
an empirical scaling $j(\chi)=j(\chi \to \infty)+a/\chi$ (inset in Fig.~\ref{fig:delta}(a)). Using the extrapolated $j$ values we have obtained the best fitted exponent $\gamma=1.01$ at $h=0.05$, 
which is within the estimated $1\%$ accuracy of the diffusive $\gamma=1$. Namely, even assuming no error due to finite $\chi$ (or extrapolation) or due to finite sample size fluctuations 
$\sigma(j)/\sqrt{M}$ when using $M$ disorder realizations, there is an ambiguity whether one should fit $\sim 1/L^\gamma$, or perhaps $\sim 1/(L-1)^\gamma$, bringing in a relative error of order 
$\sim 1/L$. The cases of $\Delta = 1.0, 0.5$ are shown in the bottom two plots of Fig.~\ref{fig:delta} .

\begin{figure}[ht!]
\centerline{\includegraphics[width=3.in]{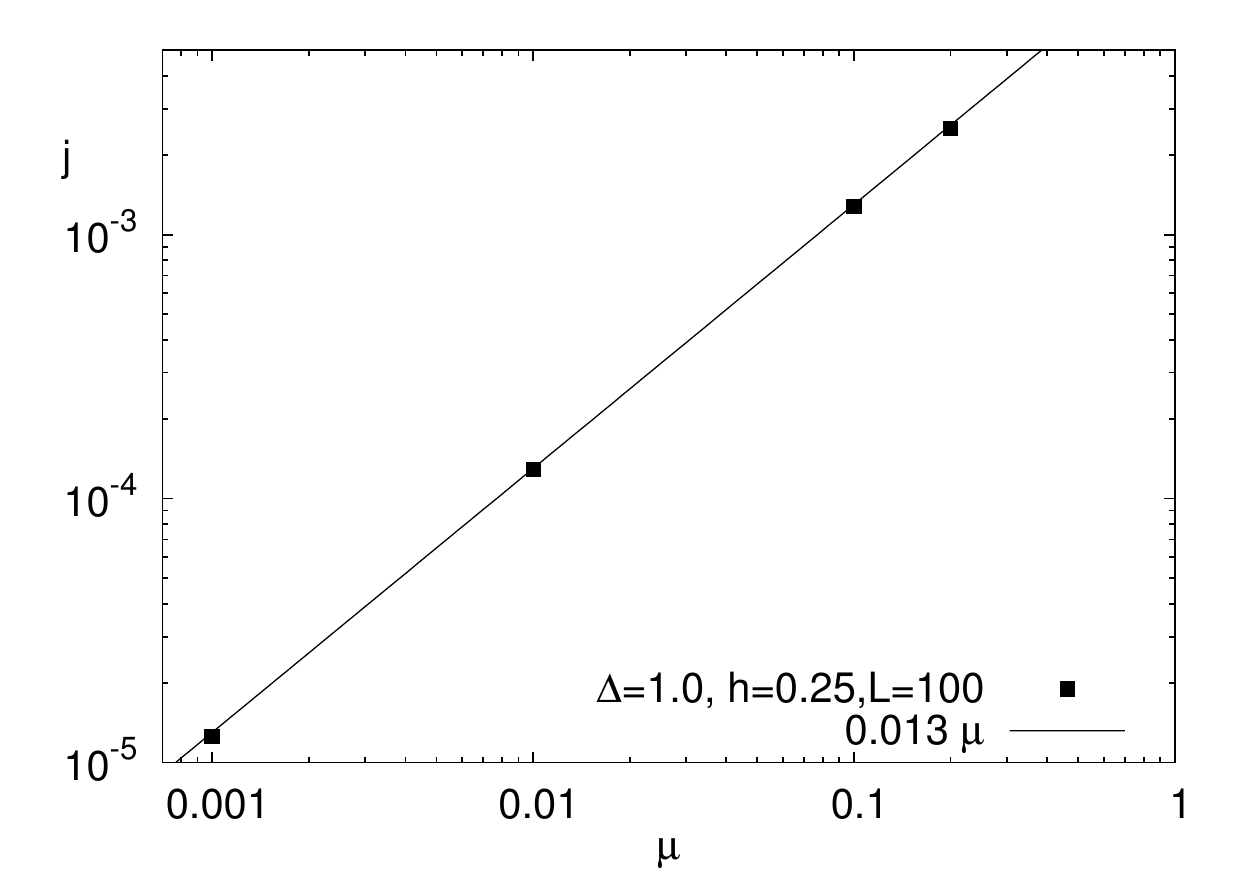}}
\caption{Dependence of average $j$ on driving strength $\mu$. One can see that the used $\mu=0.001$ is deep within (trivial, i.e., perturbative) linear dependence of $j \propto \mu$.}
\label{fig:jodmu}
\end{figure}

We also checked that at our employed driving strength $\mu=0.001$ we are deep in a regime where the spin current is linearly proportional to driving, $j \propto \mu$, see Fig.~\ref{fig:jodmu} .

\subsection{Magnetization profiles}

\begin{figure}[h!]
\centerline{\includegraphics[width=3.in]{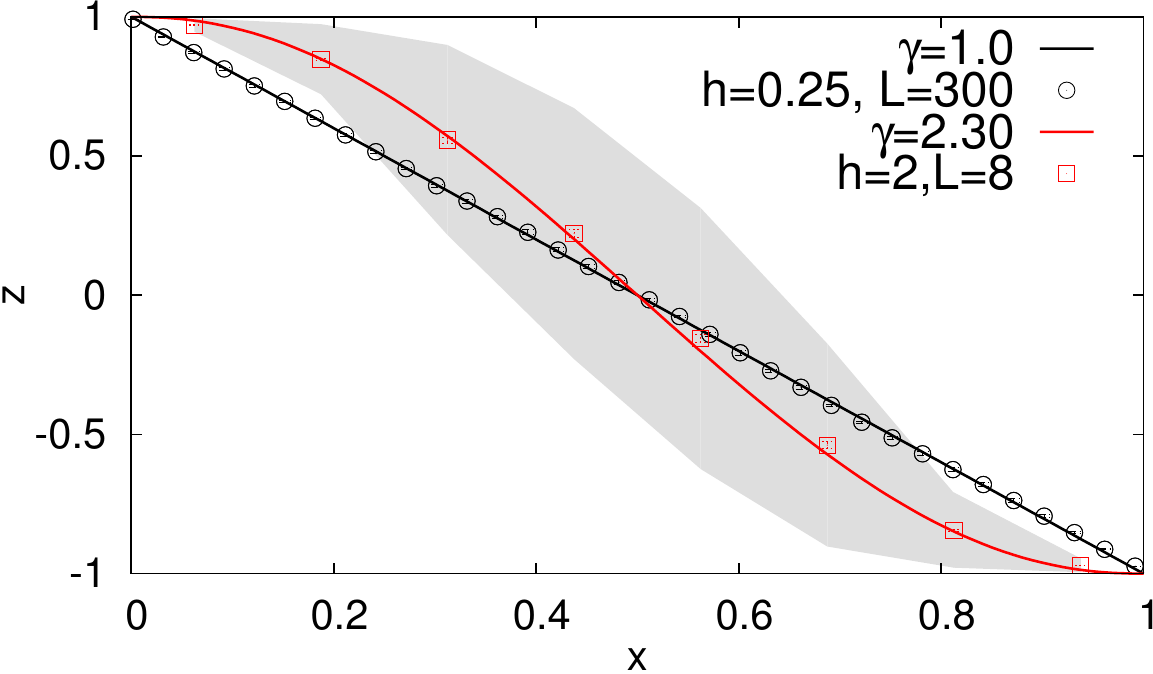}}
\caption{(Color online) Disorder averaged NESS magnetization profiles 
$z(x=\frac{k-0.5}{L})=\ave{\sz{k}}/\mu$ in the subdiffusive phase ($h=2$), 
and the diffusive one ($h=0.25$), for $\Delta=1$. Full curves are 
theoretical prediction $z(x)$ based on an empirical $\gamma$ from 
Fig.~\ref{fig:1} . Gray area for $h=2$ is standard deviation of the 
distribution.}
\label{fig:profiles}
\end{figure}

\begin{figure}[t!]
\centerline{
\includegraphics[width=1.6in]{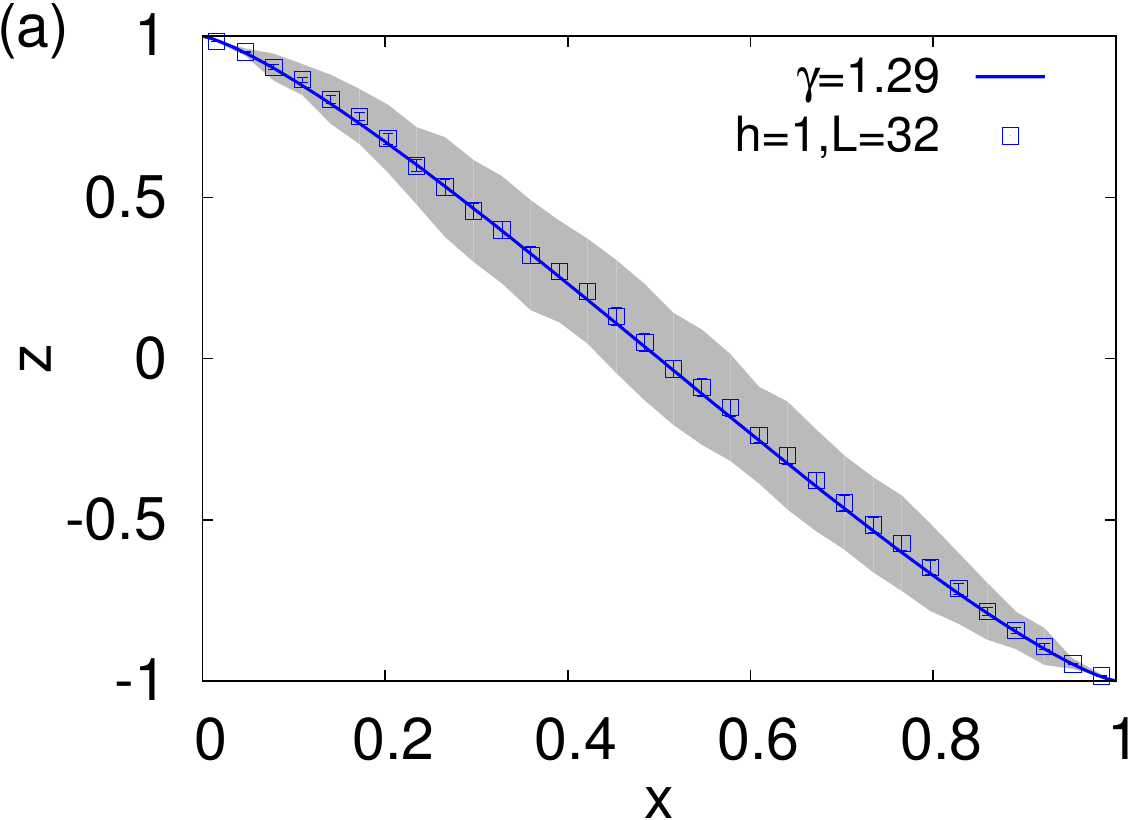}
\hskip1mm
\includegraphics[width=1.6in]{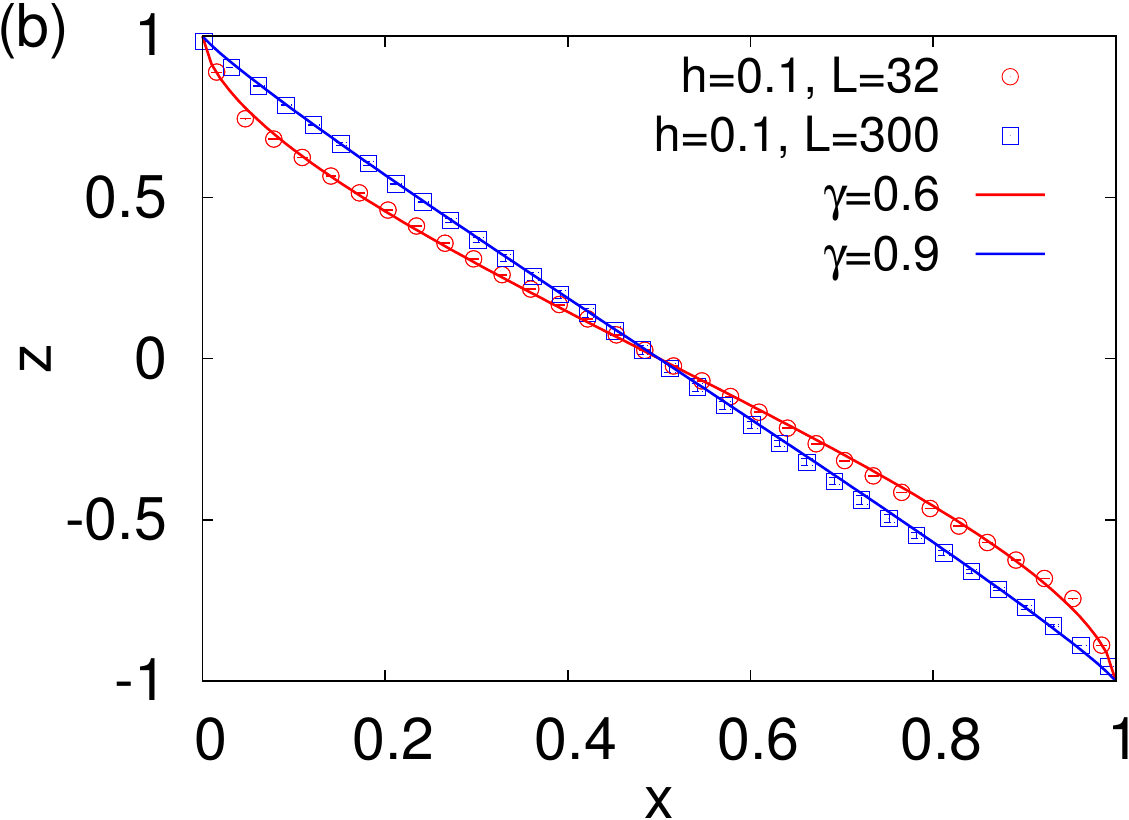}
}
\caption{(Color online) Disorder averaged NESS magnetization profiles $z(x=\frac{k-0.5}{L})=\ave{\sz{k}}/\mu$ for $\Delta=1$. 
(a) Subdiffusive phase at $h=1$. Full curve is the theory using the scaling exponent $\gamma$ from Fig.~\ref{fig:1} . (b) Diffusive phase with $h=0.1$, for which $L_*$ is larger than shown $L$. 
Theory still correctly describes profiles if one uses for $\gamma$ a local logarithmic derivative of $j(L)$. Note also how the superdiffusive signature at small $L$ gives a mirrored profile as compared 
to the subdiffusive case in the left plot.}  
\label{fig:profil}
\end{figure}

So far we have exclusively focused on the current, however, in our simulations 
we also obtain NESS magnetization profiles $\ave{s^z_k}$. We shall present a 
heuristic theory describing the observed profiles in terms of the exponent $\gamma$. For $\gamma \neq 1$ a finite-length diffusion constant $D$, defined 
via $D:=-j(L) L/(s^z_1-s^z_L)$, depends on $L$ as $D \sim L^{1-\gamma}$. This means that in a subdiffusive regime $D$ will be larger for smaller systems, 
and one can argue that in a given system close to boundaries the 
local resistivity will be effectively smaller than in the bulk, resulting 
in smaller magnetizaton gradient close to the edge. 
A heuristic simple way to account for that is to assume a 
space-dependent diffusion constant of form~\cite{JSTAT11} $D(x) \propto [x(1-x)]^{1-\gamma}$, which has correct behavior $D(1/L)/D(0.5) \sim 1/L^{1-\gamma}$ close to boundaries ($x\in [0,1]$ is a scaled spatial coordinate). Integrating the 
transport law $j=-D(x) {\rm d}z/{\rm d}x$ for a given $D(x)$ one obtains a scaled 
magnetization profile $z(x)=1-2B_\gamma(x)/B_\gamma(1)$, where $B_\gamma(x):=\int_0^x t^{\gamma-1}(1-t)^{\gamma-1}{\rm d}t$ is Beta function. In Fig.\ref{fig:profiles} and Fig.~\ref{fig:profil}(a) one can see good agreement between numerically computed profiles and the theoretical prediction. 
The agreement is good even for system sizes for which $\gamma(L)$ (local logarithmic derivative of $j(L)$ in Fig.~\ref{fig:1}) has not yet converged to its 
TDL value, as demonstrated in Fig.~\ref{fig:profil}(b). Namely, for weak disorder $h=0.1$ where $L_*$ is very large ($20/h^{1.33} \approx 430$), 
the displayed results for $L=32$ and $L=300$ are not yet in the asymptotic TDL regime. 
Still, the theoretical $B_\gamma(x)$ can be used if one takes for $\gamma$ a local slope of the curve $j(L)$ (in the log-log plot) at a given $L$. The employed values 
$\gamma=0.6$ and $\gamma=0.9$ are indeed the slopes read from Fig.~\ref{fig:1}(a). Note also how the space-dependence of the diffusion constant 
(nontrivial derivative of $z(x)$) decreases as $L \rightarrow L_*$ for this particular $h$, for which one obtains diffusion in the TDL.

We finally note that the choice of $D \propto [x(1-x)]^{1-\gamma}$ is somewhat arbitrary; for other choices one can also have correct $D(1/L)/D(0.5) \sim L^{\gamma-1}$ scaling close to boundaries. One possibility would be for instance $D(x) \propto [\sin{(\pi x)}]^{1-\gamma}$, leading to a more complicated expression for profile $\tilde{z}(x)$ in terms of Hypergeometric function, $\tilde{z}(x)={}_{2}F_1(\frac{1}{2},1-\frac{\gamma}{2};\frac{3}{2}; \cos^2{(\pi x)})\frac{2}{\sqrt{\pi^3}}\sin{(\gamma \frac{\pi}{2})} \Gamma(1-\frac{\gamma}{2})\Gamma(\frac{1+\gamma}{2})\cos{(\pi x)}$. For integer $\gamma$ the expression for $\tilde{z}(x)$ simplifies to a trigonometric polynomial of degree $(\gamma-1)$ (plus a linear function for odd $\gamma$). With our numerical precision we can not distinguish between $z(x)$ and $\tilde{z}(x)$ as the two differ in relative terms by less than $5\%$. They would differ more for either very small or very large $\gamma$.

\subsection{Energy current}

A generic Lindblad driving will induce a nonzero spin current $j_k$ as well as a nonzero energy current $j^{\rm E}_k$. Energy current is $j^{\rm E}_{k}:=\ii [H_{k-1,k},H_{k,k+1}]=\mathbf{s}^\Delta_{k-1}\cdot (\mathbf{s}_k \times \mathbf{s}^\Delta_{k+1})+\frac{h_k}{2}(j_{k-1}+j_k)$, where $\mathbf{s}^\Delta_k:=(\sx{k},\sy{k},\Delta \sz{k})$, 
and can alternatively be expressed as $j^{\rm E}_k=\Delta \sz{k-1} j_{k,k+1}-\sz{k}j_{k-1,k+1}+\Delta \sz{k+1}j_{k-1,k}+\frac{h_k}{2}(j_{k-1,k}+j_{k,k+1})$, where $j_{k,r}:=\sx{k}\sy{r}-\sy{k}\sx{r}$. 
For our choice of Lindblad ``magnetization'' driving and nonzero disorder $h_k$, for each particular disorder realization both currents are generally nonzero in the steady-state $\rho_\infty$ 
(energy current $j^{\rm E}_k$ though is typically much smaller than the spin current $j_k$). 
However, upon disorder averaging (if a disorder distribution is even under $h_k \to -h_k$) the average energy current is zero. 
Therefore, our driving on average induces only spin current. This zero average energy current is a consequence of a symmetry. 

\begin{figure}[t!]
\centerline{\includegraphics[width=2.9in]{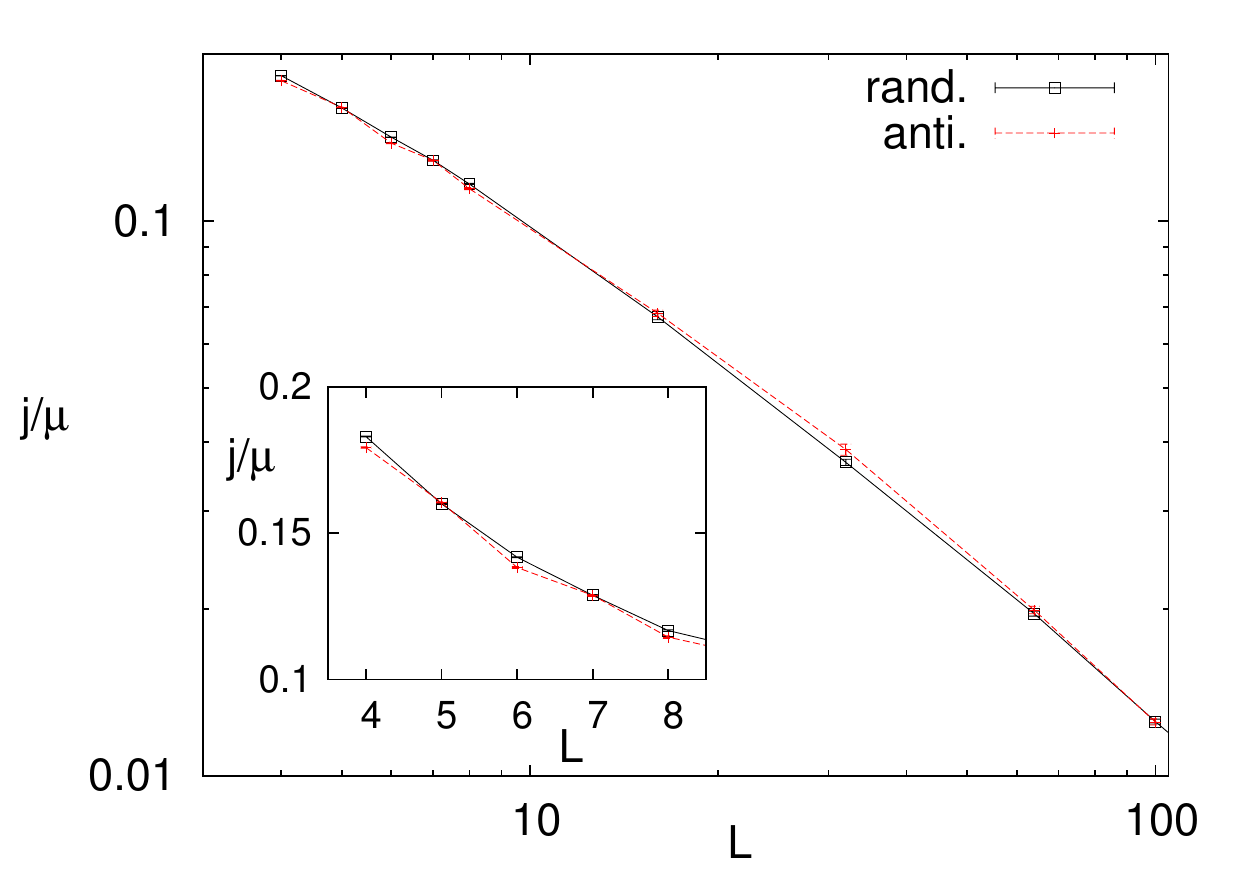}}
\caption{(Color online) Comparison between NESS currents for completely independent disorder at all sites and the one with disorder that is odd under spatial reflection, $h_k=-h_{L+1-k}$ 
(labeled ``anti.''). The two sets of points are within errorbars of each other, except for small $L<10$ visible in the inset. All data are for $\Delta=1$, $h=0.25$.}
\label{fig:anti}
\end{figure}

One could also make the steady-state energy current exactly zero even for each disorder realization separately by choosing disorder that is even under reflection $R$ across the middle of the chain, 
$h_k=-h_{L+1-k}$. Because for such antisymmetric disorder there are disorder correlations only over lengths of order $\sim L$, in the TDL physics should be the same as in our case of completely 
independent $h_k$. This is indeed what is also verified by numerics in Fig.~\ref{fig:anti} .

\begin{figure}[ttp!]
\centerline{\includegraphics[width=3.in]{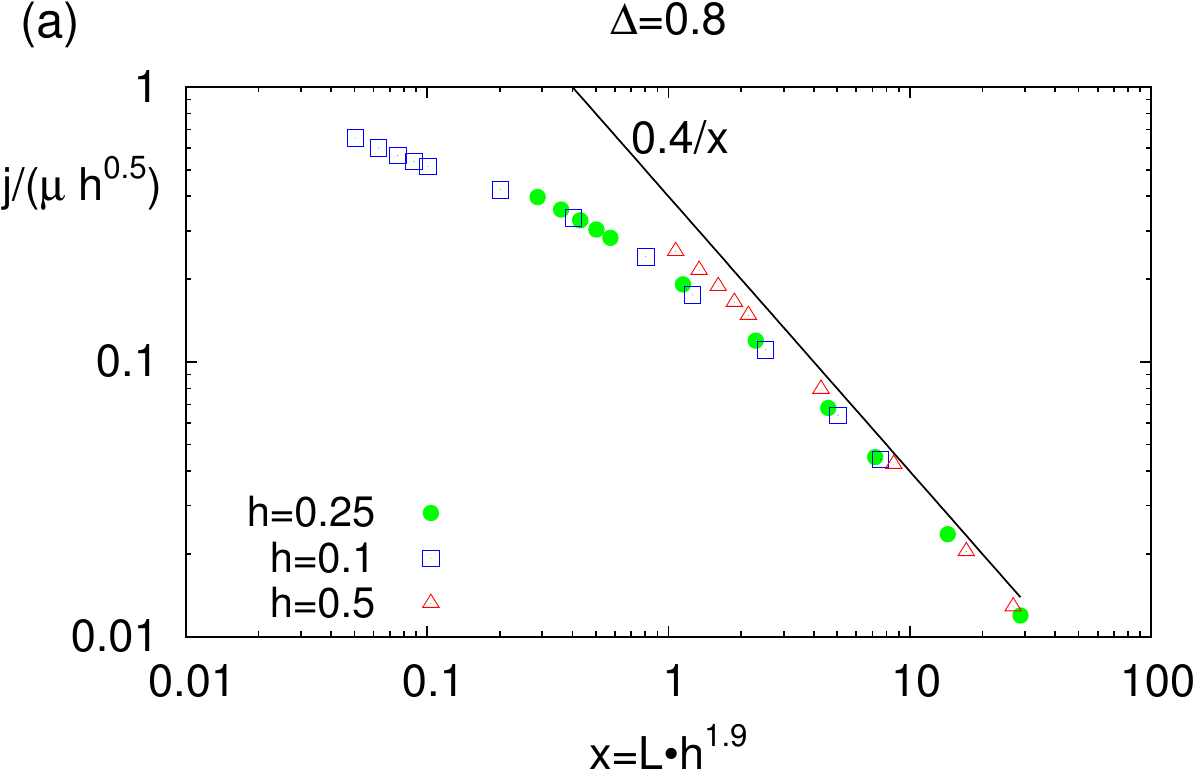}}
\vskip3mm
\centerline{\includegraphics[width=3.in]{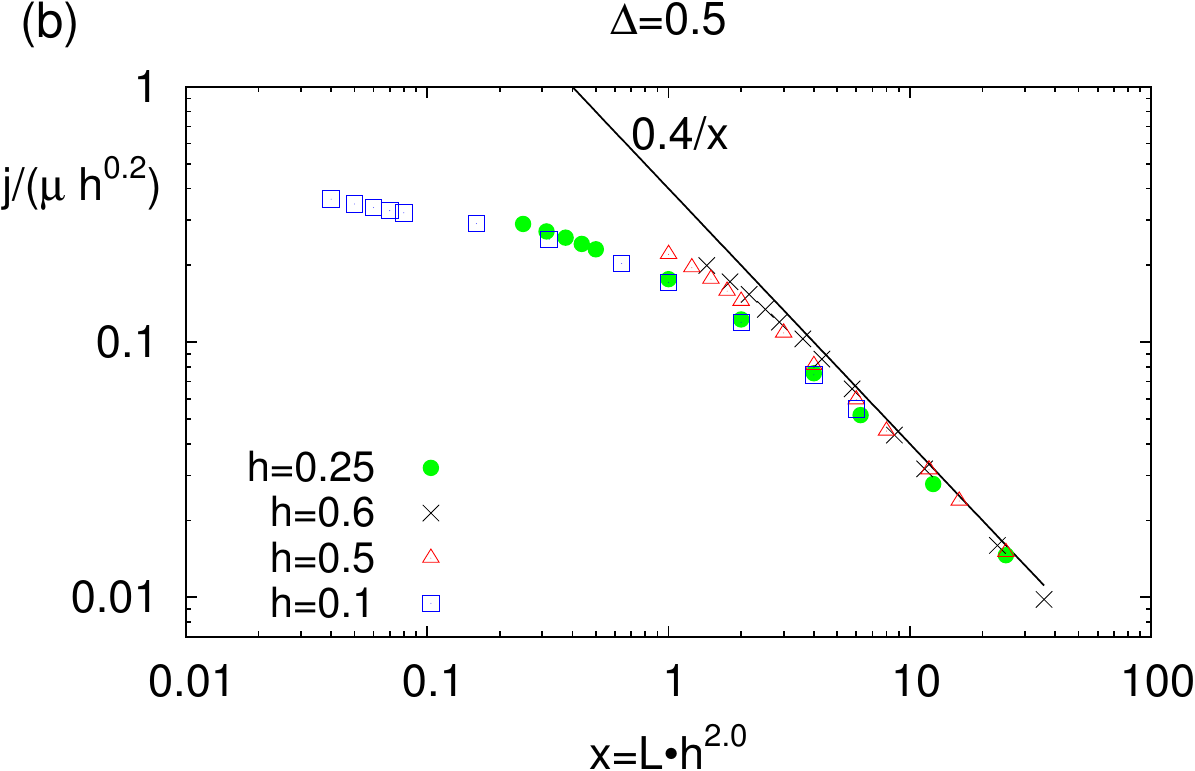}}
\centerline{\includegraphics[width=3.1in]{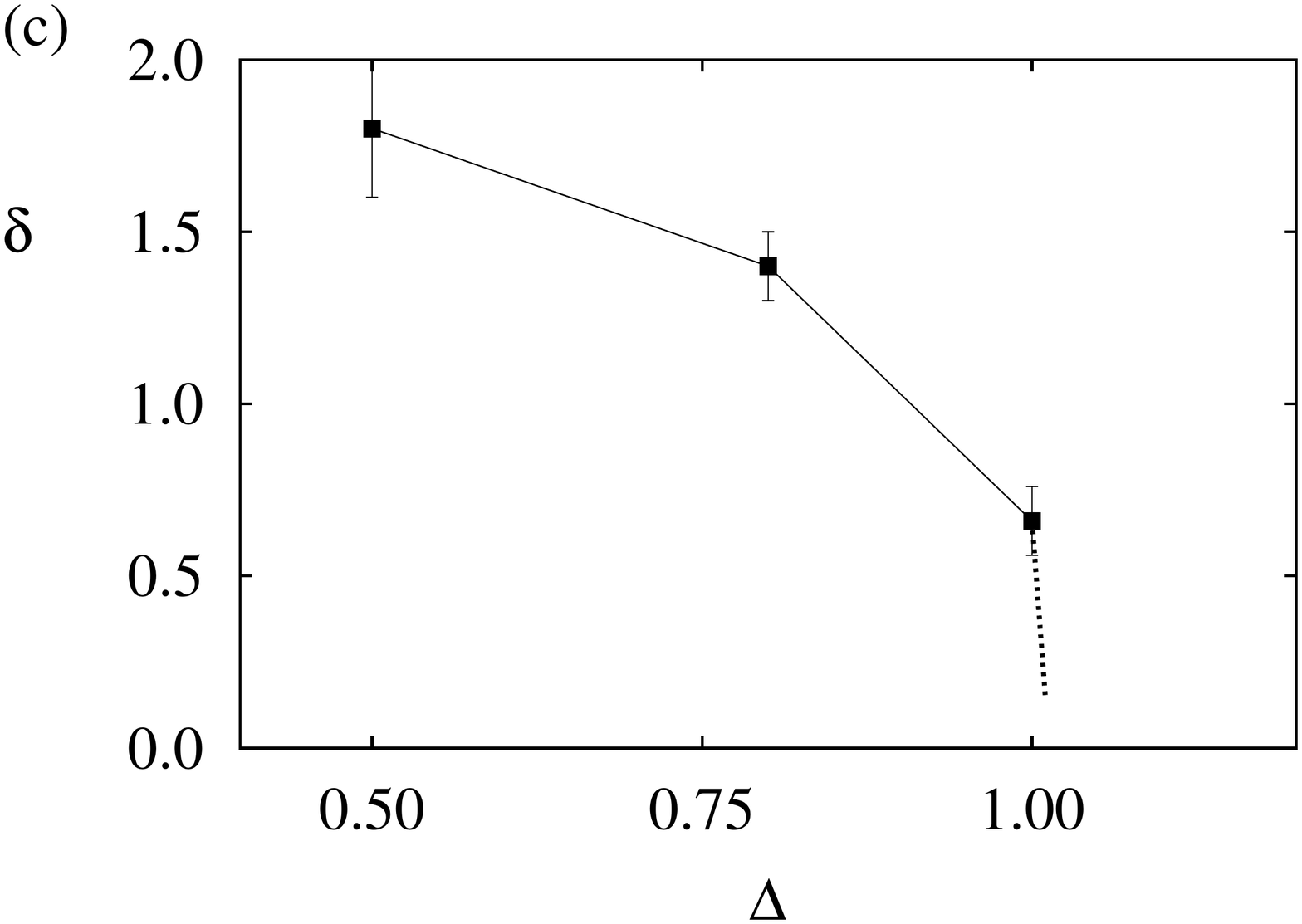}}
\vspace{-1cm}
\caption{(Color online) Scaling of $L_* \sim 1/h^\nu$ and exponent $\delta$ for small $h$. Collapse of data points is achieved when the x-axis is scaled as $Lh^\nu$ and the y-axis by $j/h^{\nu-\delta}$. 
Frame (a) shows data for $\Delta=0.8$, while (b) is for $\Delta=0.5$. System sizes $L \le 100$ for $h\ge 0.5$, $L \le 400$ for $h=0.25$ and $L \le 600$ for $h=0.1$ are shown.
Frame (c) shows exponent $\delta$ in $D \sim h^{-\delta}$ signaling scaling of diffusion constant at weak disorders as a function of anisotropy $\Delta$.
}
\label{fig:smallhD}
\end{figure}
Zero energy current in the case of an antisymmetric disorder is a consequence of energy current being odd under $S:=RP$, where $R$ is a spatial reflection of lattice sites, $k \to L+1-k$, and $P:=\prod_{k=1}^L \sigma^{x}_k$ is a parity (particle-hole) transformation. One can see that spin current is even under $S$, $S(j_k)=j_{L+1-k}$, while energy current is odd, $S(j^{E}_k)=-j^{E}_{L+1-k}$ provided $h_k=-h_{L+1-k}$. Our Lindblad equation is always invariant under $S$, meaning that the unique steady-state is also invariant, $S(\rho_\infty)=\rho_\infty$. 
As a consequence, provided $h_k=-h_{L+1-k}$, one has to have $\tr{(\rho_\infty j^{E}_k)}=0$ for each disorder realization. If the disorder is not antisymmetric for each realization separately, disorder averaging can still restore the symmetry, e.g., provided $\ave{h_k}=0$ as is in the case for a the box-distribution that we use.

\subsection{Weak disorder scaling}

In the main part we demonstrated that for small disorder in the isotropic model, $\Delta=1$, there is a length scale $L_* \sim 1/h^{1.33}$ below which one is not yet in the diffusive regime, 
and we explained this length-scale as arising due to scattering and the superdiffusive scaling relation between length and time, $x \sim t^\beta$. 

To provide additional support for such an explanation we have also checked the scaling for $\Delta<1$, where we rigorously know that the spin transport is ballistic~\cite{Prosen11}, i.e., $\beta=1$, 
which should therefore lead to scaling $L_* \sim 1/h^2$. In Figs.~\ref{fig:smallhD}(a),(b) we show a collapse of the data for $\Delta=0.8$ and $\Delta=0.5$, thereby further vindicating our theory. From the scaling of the $x$-axis we indeed see that $L_* \sim 1/h^\nu$ with $\nu$ being close to $2$. 
Divergence exponent $\delta$ of the diffusion constant at small $h$ is in turn equal to the difference of scaling exponents employed on the $x$- and $y$-axes, 
leading to $\delta\approx 1.4\pm 0.1$ at $\Delta=0.8$, and $\delta \approx 1.8\pm 0.2$ at $\Delta=0.5$ (we have also independently checked these exponents by directly fitting $j \sim D(h)/L$). 
Therefore, the divergence gets stronger as one moves away from the isotropic point. This is displayed in Fig.~\ref{fig:smallhD}(c). Dotted line indicates our expectation for the behavior of $\delta$ as 
we enter the $\Delta > 1$ regime, as we now explain.

The regime with $\Delta>1$ is different because, as opposed to clean superdiffusive transport for $\Delta \le 1$, it instead shows a diffusive scaling of the spin current with system size~\cite{PRL11} (see though Ref.~\cite{fluctXXZ} for the scaling of higher current cumulants). 
\begin{figure}[h!]
\centerline{\includegraphics[width=3.2in]{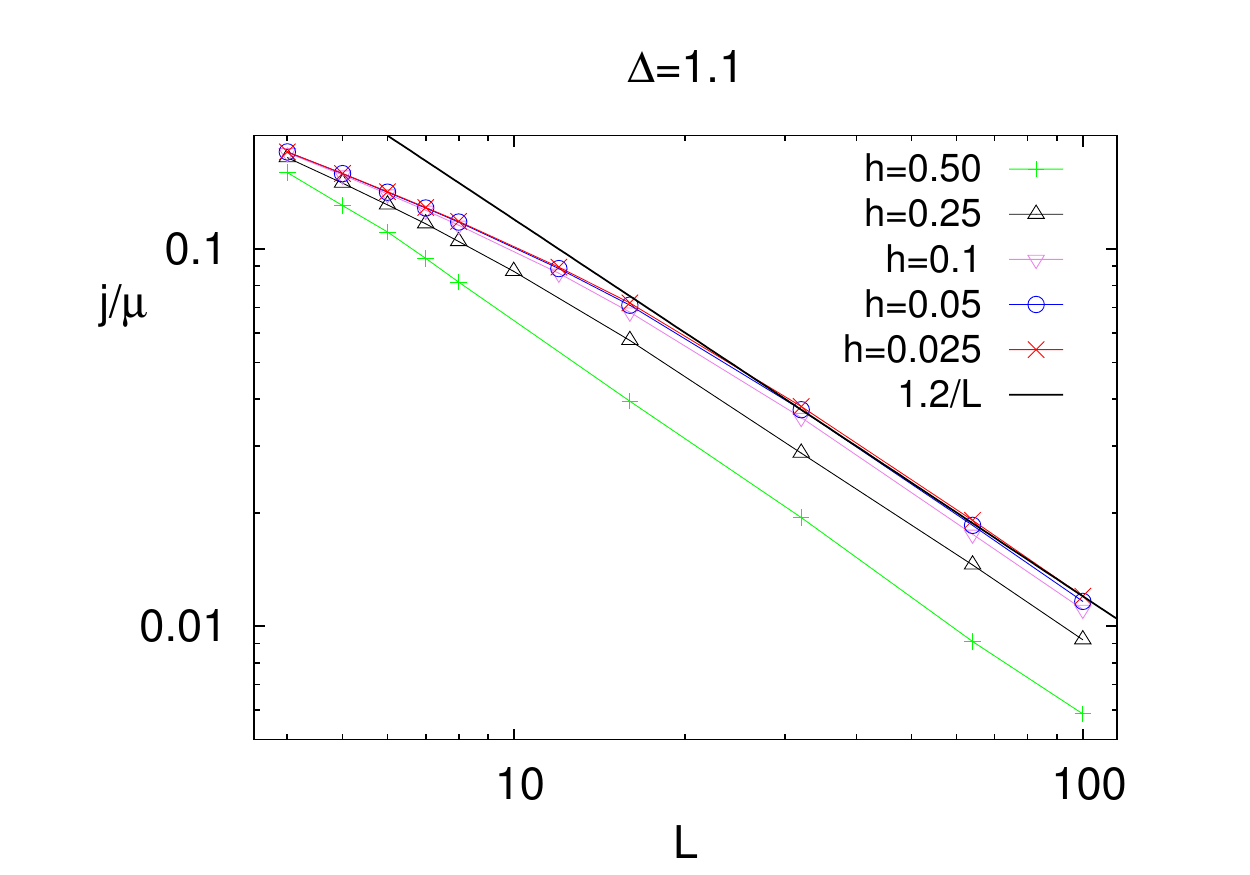}}
\caption{(Color online) Scaling of spin current for small $h$ and $\Delta=1.1$. For small $h$ current becomes independent of disorder, implying that the crossover scale $L_*$ saturates (i.e., does not diverge in the limit $h \to 0$ as for $\Delta \le 1$).}
\label{fig:smallhG}
\end{figure}
Diffusive clean conductivity sets an upper limit on transport in the presence of disorder. 
Namely, assuming that the total scattering rate is a sum of scattering rate due to disorder and a scattering rate due to interaction, for sufficiently weak disorder $h$ scattering due to interaction will prevail. 
Therefore, for small $h$ the whole dependence $j(L)$ should become independent of $h$, and with it also a crossover length-scale $L_*$, implying in turn the scaling exponents are $\delta=\nu=0$. 
Such a scenario is confirmed by numerical data in Fig.~\ref{fig:smallhG} .

\end{document}